\documentclass[fleqn,usenatbib]{mnras}

\usepackage{newtxtext,newtxmath}

\usepackage[T1]{fontenc}

\DeclareRobustCommand{\VAN}[3]{#2}
\let\VANthebibliography\thebibliography
\def\thebibliography{\DeclareRobustCommand{\VAN}[3]{##3}\VANthebibliography}

\usepackage{graphicx}	
\usepackage{amsmath}	

\title[Bargiacchi G. et al.: Quasar cosmology]{Quasar cosmology: dark energy evolution and spatial curvature}

\author[G. Bargiacchi et al.]{
G. Bargiacchi,$^{1,2}$\thanks{E-mail: giada.bargiacchi@unina.it}
M. Benetti,$^{1,2}$
S. Capozziello$^{1,2,3}$
E. Lusso$^{4,5}$
G.Risaliti$^{4,5}$
and M. Signorini$^{4,5}$
\\
$^{1}$Scuola Superiore Meridionale, Largo S. Marcellino 10, 80138 Napoli, Italy \\
$^{2}$Istituto Nazionale di Fisica Nucleare (INFN), Sez. di Napoli, Complesso Univ. Monte S. Angelo, Via Cinthia 9, 80126, Napoli, Italy\\
$^{3}$Dipartimento di Fisica "E. Pancini", Università degli Studi di Napoli Federico II, Complesso Univ. Monte S. Angelo, Via Cinthia 9, 80126, Napoli, Italy\\
$^{4}$Dipartimento di Fisica e Astronomia, Università degli Studi di Firenze, via G. Sansone 1, 50019 Sesto Fiorentino, Firenze, Italy\\
$^{5}$Istituto Nazionale di Astrofisica (INAF) – Osservatorio Astrofisico di Arcetri, 50125 Florence, Italy
}

\date{Accepted XXX. Received YYY; in original form ZZZ}

\pubyear{2022}

\begin{document}
\label{firstpage}
\pagerange{\pageref{firstpage}--\pageref{lastpage}}
\maketitle

\begin{abstract}
We analyse some open debates in cosmology in light of the most updated quasar (QSO) sample, covering a wide redshift range up to $z\sim7.5$, combined with type Ia supernovae (SNe) and baryon acoustic oscillations (BAO). Indeed, extending the cosmological analyses with high-redshift data is key to distinguishing between different cosmological models that are degenerate at low redshifts, and allowing better constraints on a possible dark energy (DE) evolution. Also, we discuss combinations of BAO, SNe, and QSO data to understand their compatibility and implications for extensions of the standard cosmological model. Specifically, we consider a flat and non-flat $\Lambda \mathrm{CDM}$ cosmology, a flat and non-flat DE model with a constant DE equation of state parameter ($w$), and four flat DE models with variable $w$, namely the Chevallier-Polarski-Linder and Jassal-Bagla-Padmanabhan models, and an ``exponential'' and Barboza-Alcaniz parameterisations. We find that a joint analysis of QSO+SNe with BAO is only possible in the context of a flat Universe. Indeed BAO confirms the flatness condition assuming a curved geometry, whilst SNe+QSO show evidence of a closed space. We also find $\Omega_{M,0}=0.3$ in all data sets assuming a flat $\Lambda \mathrm{CDM}$ model. Yet, all the other models show a statistically significant deviation at 2-3$\sigma$ with the combined SNe+QSO+BAO data set. In the models where DE density evolves with time, SNe+QSO+BAO data always prefer $\Omega_{M,0}>0.3$, $w_{0}<-1$ and $w_{a}>0$. This DE phantom behaviour is mainly driven by SNe+QSO, while BAO are closer to the flat $\Lambda \mathrm{CDM}$ model.
\end{abstract}




\section{Introduction}
\label{intro}
The most widely adopted parameterisation of the observed Universe is based on the so-called $\Lambda$ cold dark matter ($\Lambda$CDM) model \citep{peebles1984}, relying on the existence of cold dark matter and dark
energy ($\Lambda$) associated with a cosmological constant \citep{2001LRR.....4....1C} in a spatially flat geometry.
Predictions from this model have been found to agree with most of the observational probes such as the cosmic microwave background \citep[CMB; e.g.][]{planck2018}, the baryon acoustic oscillations \citep[BAO; e.g.][]{eboss2021}, and the present accelerated expansion of the Hubble flow, based on the distance modulus--redshift relation (the so-called Hubble-Lema\^itre, or simply Hubble diagram) of type Ia supernovae \citep[SNe Ia; e.g.][]{riess1998,perlmutter1999}, where a dominant dynamical contribution, dubbed dark energy (DE) and related to the cosmological constant, should drive such an acceleration. 
However, the fundamental physical origin and the properties of DE are still unknown, as the interpretation of $\Lambda$ is plagued by a severe  fine-tuning issue to obtain the right amount of DE observed today. Moreover, the data sets listed above do not fully fit the evolution of  DE ranging from early to late epochs \citep{Benetti:2019lxu,2021JCAP...10..008Y} and do not fully rule out  a spatially non-flat Universe \citep{Park:2017xbl,2020NatAs...4..196D,Handley:2019tkm,DiValentino:2020hov,2021JCAP...10..008Y}. The latter possibility has raised a remarkable debate about the importance of properly combining CMB data to infer significant statistical interpretations from the analysis \citep{planck2018,Efstathiou:2020wem} and, by extension, the importance of combining data sets that do not reveal manifest tension \citep{2021ApJ...908...84V,2021JCAP...11..060G}.
Deviations from the spatially flat $\Lambda$CDM model would imply important theoretical and observational consequences and a change in our current understanding of cosmic evolution \citep[e.g.][]{CapoSpa}. Statistically significant deviations in this directions have already been found in cosmological analyses with high-redshift probes such as Gamma-Ray Bursts (see \citealt{Dainotti2008,dainotti11a,Dainotti11b,Dainotti2013a,Dainotti2013b,dainotti15,dainotti17a,Dainotti2020a,Dainotti2020b} for the standardisation of these sources as cosmological candles) and quasars (QSOs) combined with SNe Ia \citep{rl19,lusso2019,2020A&A...642A.150L,refId0}. Such a joint analysis (SNe+QSO)
makes use of the observed non-linear relation between the ultraviolet and the X-ray luminosity in QSOs \citep[e.g.][]{steffen06,just07,2010A&A...512A..34L,lr16,2021A&A...655A.109B,2022ApJ...931..106D} to provide an independent measurement of their distance \citep[see e.g.][for details]{rl15,rl19,2020A&A...642A.150L}. The methodology is complementary to the traditional resort to type Ia SNe to estimate the cosmological parameters, yet it extends the Hubble-Lema\^itre diagram to a redshift range currently inaccessible to SNe ($z=2.4-7.5$).
Within a model where an evolution of the DE equation of state (EoS) in form $w(z)=w_0+w_a\times z/(1+z)$ is assumed, the data suggest that the DE parameter is increasing with time \citep{rl19,2020A&A...642A.150L}. Therefore, it is compelling to further study extensions of the $\Lambda$CDM model that could produce such behaviour of DE.

Here we further analyse the most updated QSO and SNe Ia data sets, in combination with BAO, to constrain cosmological parameters in a suit of cosmological models that extend the flat $\Lambda$CDM. In particular, data from BAO proved to agree with the description of a geometrically flat Universe \citep{2021JCAP...11..060G,eboss2021}, and it is important to investigate whether can be correctly combined with QSOs that give different indications. We therefore conduct a detailed analysis of the compatibility of the data, considering different DE parameterisations.
Specifically, we consider the standard cosmological model, where  DE is described by the constant $\Lambda$, and the simple extension where the DE density parameter is allowed to vary even if its EoS is constant in time ($w$CDM model). 
We study these models both in the context of a spatially flat  and a  curved Universe.
In addition, we extend our analysis by also considering, under the flatness assumption, four parametric descriptions where the DE EoS is evolving with redshift, namely the Chevallier-Polarski-Linder \citep[CPL; e.g.][]{CHEVALLIER_2001,2003PhRvL..90i1301L} and Jassal-Bagla-Padmanabhan \citep[JBP; e.g.][]{2005MNRAS.356L..11J} models, an ``exponential'' \citep[see][]{2019PhRvD..99d3543Y}, and the Barboza-Alcaniz (BA; \citealt[e.g.][]{2008PhLB..666..415B}) parameterisations.

This work shows manifestly important points of originality with respect to some previous similar analyses \citep[e.g.][]{2020ApJ...900...70R,2020MNRAS.497..263K,2021SCPMA..6459511Z}. In particular, we make use of the most updated QSO sample \citep{2020A&A...642A.150L} that has been specifically selected for cosmological studies and significantly improved compared to the previous one concerning the number of sources, quality of measurements and redshift range extension. Moreover, we also focus on and consider the compatibility among the different probes thoroughly, which is not taken into account in other works in a comprehensive manner and instead plays a crucial role in the discussion and in the statistical credibility of the results.

The manuscript is organised as follows. Section~\ref{Data sets} describes the data sets used in this work and the methods applied to analyse them. In Sect.~\ref{DE EoS parameterisations}, we introduce different (spatially flat and non-flat) cosmological models also in the context of an evolving DE, whilst in Sect.~\ref{Cosmological results}
we present the main results of the fits of these models on the SNe Ia, QSO, and BAO data sets comparing them and discussing the implications. Finally, in Sect.~\ref{Conclusions}, we summarise our findings. In Appendix \ref{appendixcpl} we describe the issues faced when applying our fitting technique to some of the models studied. 

\section{Data sets}
\label{Data sets}

Here, we work with large-scale data, selecting and combining complete samples of QSO, SNe Ia, and BAO measurements to investigate the late-time Universe. In this section, we describe each data set and the methodology used for the cosmological analyses. These are carried out using the probes separately, followed by the joint analysis. The extension to early-time CMB data will be included in a forthcoming paper that will extend the present results.

\subsection{Supernovae Ia}
For SNe Ia, we consider the most recent collection of 1048 sources from the \textit{Pantheon} sample \citep{scolnic2018}. We directly use the values of the distance moduli (with their uncertainties) to get consistency with the physical quantities of QSOs and be able to join these two probes. Indeed, as we explain in the next section, we need to calibrate QSO distances making use of SNe. To this aim, we assume an absolute magnitude $M=-19.36$ and we obtain the distance moduli by computing $m-M$, where $m$ is the apparent magnitude from the \textit{Pantheon} data.

All the fits presented in this work are obtained through
the Python package emcee \citep{2013PASP..125..306F}, which
is a pure-Python implementation of Goodman \& Weare’s affine invariant Markov chain Monte Carlo (MCMC) ensemble sampler. The likelihood function ($LF$) used in this method to fit the SNe sample, as all measurements are independent, is defined as\footnote{For the sake of simplicity we always use $\text{ln}$ instead of $\text{log}_{\mathrm{e}}$ and $\text{log}$ instead of $\text{log}_{10}$.}
\begin{equation} \label{lfsne}
\text{ln}(LF)_{\text{SNe}} = -\frac{1}{2} \sum_{i=1}^{N} \left[ \frac{(y_{i}-\phi_{i})^{2}}{\sigma^{2}_{i}} + \text{ln}(\sigma^{2}_{i})\right]
\end{equation}
where $N$ is the number of sources, $y_{i}$ is the distance modulus with statistical uncertainty $\sigma_{i}$ and $\phi_{i}$ is the distance modulus predicted by the cosmological model considered, yet it depends on the redshift and the free parameters of the model. As shown in Eq. \eqref{lfsne}, we make use of statistical uncertainties only, without considering the complete non-diagonal covariance matrix that includes also the systematic ones. Our main results do not depend on this choice, as shown in Appendix \ref{appendixsys}. Specifically, all free parameters fitted in each of the models studied in this work are compatible within 1$\sigma$ if we compare the SNe (only stat) +QSO and the SNe (stat+sys) + QSO samples. A couple of examples are shown in Appendix \ref{appendixsys}.

\subsection{Quasars}
The sample of QSOs used in this work is the one described in \citet{2020A&A...642A.150L} with the cutting at redshift $z>0.7$ for the sources with a photometric determination of the UV flux (see the mentioned paper for an explanation of this selection filter). This is composed of 2036 sources that cover the redshift range up to $z =7.54$ \citep{banados2018}. These sources have been carefully selected for cosmological studies and we refer to \citet{rl15}, \citet{lr16}, \citet{rl19}, \citet{salvestrini2019} and \citet{2020A&A...642A.150L} for a detailed description and validation of the procedure used to turn them into standard candles and for an explanation of the fitting technique used to include them in the cosmological analysis. Here, we only summarise the crucial points required by the present work.

As anticipated in the Introduction, the method to determine QSO distances is based on the non-linear relation between their UV and X-ray luminosity \citep{steffen06,just07,2010A&A...512A..34L,lr16,2021A&A...655A.109B,2022ApJ...931..106D}. The fitted distance moduli\footnote{Throughout the paper the abbreviation ``DM'' used in the formulae stands for the distance modulus and not for the dark matter component.} are obtained from $\text{DM}(z) = 5 \text{log}[D_\mathrm{{L}}(z)(\mathrm{Mpc})] + 25$ where the luminosity distance $D_\mathrm{{L}}(z)$ comes from \citep{rl15,2020A&A...642A.150L}
\begin{equation} \label{dlqso}
\text{log}D_\mathrm{{L}}(z) = \frac{\left[\text{log}F_{X} - \beta -\gamma\, (\text{log}F_{UV}+27.5)\right]}{2(\gamma -1)} - \frac{1}{2}\text{log}( 4 \pi) +28.5.
\end{equation}
In this expression, $D_\mathrm{{L}}$ is expressed in units of cm and is normalised to 28.5 in logarithm, $F_{X}$ and $F_{UV}$ are the measured flux densities (in $\mathrm{erg \, s^{-1} \, cm^{-2} \, Hz^{-1}}$) at 2 keV and 2500 \AA, respectively, and $F_{UV}$ is normalised to the logarithmic value of -27.5. The slope $\gamma$ and the intercept $\beta$ of the logarithmic X-UV luminosity relation are free parameters of the fit. The $LF$ used by the MCMC process to fit QSOs is
\begin{equation} \label{lfqso}
\text{ln}(LF)_{\text{QSO}} = -\frac{1}{2} \sum_{i=1}^{N} \left[ \frac{(y_{i}-\phi_{i})^{2}}{s^{2}_{i}} + \text{ln}(s^{2}_{i})\right].
\end{equation}
In this case, the data $y_{i}$ correspond to $\text{log}F_{X}$, while $\phi_{i}$ to the logarithmic X-flux predicted by the model assumed through the inversion of Eq.\eqref{dlqso} and it depends on $\gamma$, $\beta$, the redshift and the specific parameters of the model. Moreover, $s^{2}_{i} = dy^{2}_{i} + \gamma^{2} dx^{2}_{i} + \delta^{2}$ and it takes into account the statistical uncertainties on $\text{log}F_{X}$ ($y$) and $\text{log}F_{UV}$ ($x$), but also the intrinsic dispersion $\delta$ of the X-UV flux relation, which is another free parameter of the fit. Practically, $LF_{\text{QSO}}$ is just the same likelihood function used for SNe (Eq. \eqref{lfsne}), but modified to include the contribution of the intrinsic dispersion of the X-UV relation.

Besides, we also need an additional free parameter ($k$) to calibrate QSO distances using the distance ladder through SNe. Indeed, QSOs do not provide absolute distances as they are not able to fix the ``zero-point'' of the Hubble diagram. This is the reason why we include in our analysis the parameter $k$ shared by both SNe and QSOs: it is a rigid shift of the QSO Hubble diagram to match the one of SNe in the common redshift range. We require the distance moduli predicted by the cosmological model to be $\text{DM}(z) = 5 \text{log}[D_\mathrm{{L}}(z)(\mathrm{Mpc})] + 25 + k$ for both SNe and QSOs. This way this parameter is also degenerate with the Hubble constant $H_{0}$ that appears in the cosmological expression for $D_\mathrm{{L}}(z)$ (see Eq. \eqref{Dl(z)}). For this reason, we always assume $H_{0}$ fixed to the arbitrary value $H_{0} = 70\,  \mathrm{km} \, \mathrm{s^{-1}} \,\mathrm{Mpc^{-1}}$, while $k$ is fitted and represents the deviation of $H_{0}$ from the chosen reference value. As we have just stressed, the Hubble diagram of QSOs cannot be built without the inclusion of SNe, that act as calibrators, thus the joint likelihood function used for the combined sample of QSOs and SNe is given by $\text{ln}(LF)_{\text{QSO+SNe}} = \text{ln}(LF)_{\text{SNe}}+\text{ln}(LF)_{\text{QSO}}$.

\subsection{Baryon Acoustic Oscillations}

The BAO data we consider in the present work come from the survey 6dFGS \citep{2011MNRAS.416.3017B}, SDSS DR7 (MGS) \citep{2015MNRAS.449..835R}, BOSS DR12 \citep{2017MNRAS.470.2617A} and eBOSS \citep{2018MNRAS.473.4773A}. This is the most common data set used in recent studies and \citet{planck2018}, with the inclusion of the measurement at $z=1.52$ \citep{2018MNRAS.473.4773A}. We do not consider Ly$\alpha$ BAO determinations as they are more complicated and require additional assumptions than galaxy BAO measurement, as explained in \citet{planck2018}. We describe all the measurements used and their details in Table \ref{tableBAO}.

\begin{table}
\caption{BAO data set: survey, effective redshift $z$, physical quantity, measurement and, where necessary, fiducial comoving sound horizon at drag epoch ($r_{s,fid}$). The quantities $\displaystyle D_{V}(z) \frac{r_{s,fid}}{r_{s}(z_{d})}$,  $\displaystyle D_{M}(z) \frac{r_{s,fid}}{r_{s}(z_{d})}$ and $r_{s,fid}$ are expressed in units of Mpc, while $\displaystyle H(z) \frac{r_{s}(z_{d})}{r_{s,fid}}$ in $\mathrm{km \,s^{-1} \, Mpc^{-1}}$ and $\displaystyle \frac{r_{s}(z_{d})}{D_{V}(z)}$ is dimensionless. (1)~ \citet{2011MNRAS.416.3017B}; (2) \citet{2015MNRAS.449..835R}; (3) \citet{2017MNRAS.470.2617A}; (4) \citet{2018MNRAS.473.4773A}.}
\label{tableBAO}
\centering 
\setlength{\tabcolsep}{0pt}
\begin{tabular}{ c c c c c c} 
\hline
Survey & $z$ & Quantity & Measurement & $r_{s,fid}$ & Ref.\\
\hline
6dFGS & 0.106 & $\displaystyle \frac{r_{s}(z_{d})}{D_{V}(z)}$ & 0.336 $\pm$ 0.015 & \ & 1 \\
SDSS DR7(MGS) & 0.15 & $\displaystyle D_{V}(z) \frac{r_{s,fid}}{r_{s}(z_{d})}$ & 664 $\pm$ 25 & 148.69 & 2 \\
BOSS DR12 & 0.38 & $\displaystyle D_{M}(z) \frac{r_{s,fid}}{r_{s}(z_{d})}$ & 1512.39 & 147.78 & 3\\
\ &  \ & $\displaystyle H(z) \frac{r_{s}(z_{d})}{r_{s,fid}}$ & 81.2087  & 147.78 & 3\\
BOSS DR12 & 0.51 & $\displaystyle D_{M}(z) \frac{r_{s,fid}}{r_{s}(z_{d})}$ & 1975.22 & 147.78 & 3\\
\ & \ & $\displaystyle H(z) \frac{r_{s}(z_{d})}{r_{s,fid}}$ & 90.9029  & 147.78 & 3\\
BOSS DR12 & 0.61 & $\displaystyle D_{M}(z) \frac{r_{s,fid}}{r_{s}(z_{d})}$ & 2306.68 & 147.78 & 3\\
\ & \ & $\displaystyle H(z) \frac{r_{s}(z_{d})}{r_{s,fid}}$ & 98.9647  & 147.78 & 3\\
eBOSS & 1.52 & $\displaystyle D_{V}(z) \frac{r_{s,fid}}{r_{s}(z_{d})}$ & 3843 $\pm$ 147 & 147.78 & 4\\
 \hline
\end{tabular}
\end{table}

The physical quantities appearing in Table \ref{tableBAO} are the volume-averaged distance $D_{V}(z)$, the comoving sound horizon at baryon drag epoch $r_{s}(z_{d})$ and its value in the fiducial cosmology of the specific survey $r_{s,fid}$. They are defined as follows \citep[see e.g.][]{1998ApJ...496..605E,2005ApJ...633..560E}:
\begin{equation} \label{baoquantities1}
D_{V}(z) = \left[ \frac{cz}{H(z)} \frac{D_\mathrm{{L}}^{2}(z)}{(1+z)^{2}} \right]^{\frac{1}{3}}
\end{equation}
and
\begin{equation} \label{baoquantities2}
r_{s}(z_{d}) = \int_{z_{d}}^{\infty} \frac{c_{s}(z')}{H(z')}dz'
\end{equation}
where the sound speed $c_{s}$ is given by
\begin{equation} \label{cs}
c_{s}(z) =  \frac{c}{\sqrt{3 \left[ 1+ \frac{3 \Omega_{b,0}}{4 \Omega_{\gamma,0}} \frac{1}{1+z} \right]}}
\end{equation}
and the redshift of the baryon drag epoch $z_{d}$ is 
\begin{equation}\label{zd}
z_{d} = \frac{1291 (\Omega_{M,0}h^{2})^{0.251}}{1+0.659(\Omega_{M,0}h^{2})^{0.828}} \left[ 1+ b_{1}(\Omega_{b,0}h^{2})^{b_{2}} \right]  
\end{equation}
with 
\begin{align}
    b_{1} &= 0.313 (\Omega_{M,0}h^{2})^{-0.419} \left[ 1+0.607 (\Omega_{M,0}h^{2})^{0.674} \right] \\
    b_{2} &= 0.238 (\Omega_{M,0}h^{2})^{0.223}
\end{align}
The quantity $D_{M}(z)$ is simply related to the luminosity distance as $\displaystyle D_{M}(z) = D_\mathrm{{L}}(z)/(1+z)$. In all the previous formulae, we used the standard notation: $c$ is the speed of light, $H(z)$ is the Hubble parameter, $h$ is the dimensionless Hubble constant $\displaystyle h= {H_{0}}/{100 \, \mathrm{km \,s^{-1} \, Mpc^{-1}}}$, $\Omega_{b,0}$, $\Omega_{\gamma,0}$ and $\Omega_{M,0}$ are the density parameters of baryon, photon and matter component, respectively.

We note that $\Omega_{M,0}$ is a free parameter of the fit, while $D_\mathrm{{L}}(z)$ and $H(z)$ are obtained from the cosmological model assumed and depend on the specific cosmological parameters, while the other quantities are fixed: $\Omega_{b,0}h^{2}=0.0224$ (in agreement with \citealt{Hinshaw_2013} and \citealt{planck2018}) and $\Omega_{\gamma,0}h^{2}=2.469 \, \mathrm{x}\, 10^{-5}$. The way we express $D_\mathrm{{L}}(z)$ and $H(z)$ in the procedure to fit BAO is straightforward. Indeed, $H(z)$ can be easily written down as a function of the cosmological parameters of the model considered (see Eq. \eqref{E(z)}). On the other hand, in our fitting technique, we assume a cosmological model and we compute the predicted distance modulus, that is the physical quantity needed to fit standard candles as SNe and QSOs. As a consequence, we can make use of the distance modulus predicted to derive $D_\mathrm{{L}}(z)$. Finally, having both $D_\mathrm{{L}}(z)$ and $H(z)$ as a function of the cosmological free parameters, we can reconstruct and fit all the BAO quantities reported in Table \ref{tableBAO}. Considering the way we use $H_0$ for SNe and QSOs (i.e. fixing $H_{0}$ and including a degenerate parameter $k$ for cross-calibration), we decided to follow the same approach also for BAO, to get consistency among all the probes. Specifically, we include in the analysis of BAO a free parameter $k_{1}$ and we fix once again $H_{0} = 70\,  \mathrm{km} \, \mathrm{s^{-1}} \,\mathrm{Mpc^{-1}}$, using $k_1$ both in $\text{DM}(z) = 5 \text{log}[D_\mathrm{{L}}(z)(\mathrm{Mpc})] + 25 + k_{1}$ and as a correction to the fixed $H_0$ in each formula requiring the Hubble constant.

Concerning the likelihood function associated with BAO, we need to consider the data from the survey BOSS DR12 separately. Indeed, as we can note from Table \ref{tableBAO}, all the other measurements are reported with the corresponding uncertainties, as they are uncorrelated (without any volume overlap), while for the BOSS DR12 measurements the uncertainties are not reported due to the presence of correlation among them (they have been computed in three partially overlapping redshift slices). As a consequence, for all the data, except the ones from BOSS DR12, the likelihood function ($LF_{\mathrm{BAO_{1}}}$) has the same expression as the one of SNe (Eq.\eqref{lfsne})
, but in this case, $y_{i}$ (with uncertainty $\sigma_{i}$) is the data corresponding to the physical quantity reported in Table \ref{tableBAO} for the surveys 6dFGS, SDSS DR7 and eBOSS and $\phi_{i}$ is the associated modelled quantity. Instead, for BOSS DR12 measurements we take into account the correlation using
\begin{equation} \label{lfbao2}
\text{ln}(LF)_{\mathrm{BAO_{2}}} = -\frac{1}{2} \Bigg[\left(\bmath{y}-\bmath{\phi}\right)^{T} \, \mathbfss{C}^{-1} \, \left(\bmath{y}-\bmath{\phi}\right)\Bigg]
\end{equation}
where $\bmath{y}$ is the vector of the data pair $\displaystyle \left(D_{M}(z) \frac{r_{s,fid}}{r_{s}(z_{d})},H(z) \frac{r_{s}(z_{d})}{r_{s,fid}}\right)$ for every effective redshift, $\mathbfss{C}$ is the associated 6 x 6 covariance matrix \citep[see][]{2017MNRAS.470.2617A} and $\bmath{\phi}$ is the vector of modelled quantities corresponding to $\bmath{y}$. In the end, the complete likelihood function for the BAO sample is $\text{ln}(LF)_{\text{BAO}} =\text{ln}(LF)_{\mathrm{BAO_{1}}}+\text{ln}(LF)_{\mathrm{BAO_{2}}}$, while for SNe+QSO+BAO is $\text{ln}(LF)_{\text{QSO+SNe+BAO}} = \text{ln}(LF)_{\text{SNe}}+\text{ln}(LF)_{\text{QSO}}+\text{ln}(LF)_{\mathrm{BAO_{1}}}+\text{ln}(LF)_{\mathrm{BAO_{2}}}$.

\section{Cosmological models and DE EoS parameterisations}
\label{DE EoS parameterisations}


Let us now describe the models we are going to analyse in the next section with the data sets introduced above. First of all, the standard cosmological model, where the EoS of the cosmological constant is defined as $w(z) = P_{\Lambda}/\rho_{\Lambda}=-1$, with $P_{\Lambda}$ and $\rho_{\Lambda}$ the pressure and energy density of DE, respectively. In addition to DE, the other cosmological fluids considered in the model are the
non-relativistic matter component (indicated by the subscript $_M$), including both baryons ($_b$) and cold dark matter ($_{CDM}$), and the relativistic component ($_r$), composed by radiation ($_{\gamma}$) and neutrinos ($_{\nu}$). The last one makes a negligible contribution in the late Universe, but it is needed in the computation of the integral in Eq.\eqref{baoquantities2}, that is computed at high redshifts\footnote{As $z_{d} \sim 1020$ assuming a flat $\Lambda \mathrm{CDM}$ model with $\Omega_{M,0}=0.3$ and $H_{0} = 70\,  \mathrm{km} \, \mathrm{s^{-1}} \,\mathrm{Mpc^{-1}}$.}. For this reason, we include the current relativistic density parameter $\Omega_{r,0}$ in our computation of the Hubble parameter evolution, $H(z)$, fixing $\Omega_{r,0}= \Omega_{\gamma,0}+\Omega_{\nu,0} =9 \, \mathrm{x}\, 10^{-5}$. We can write the background evolution of the standard cosmological model as 
\begin{equation}
\label{lcdm}
E(z) = \frac{H(z)}{H_0} = \Bigg[\Omega_{M,0}\,(1+z)^{3} + \Omega_{r,0}\,(1+z)^{4} +
\Omega_{\Lambda,0}\Bigg]^{\frac{1}{2}}\,
\end{equation}
where the only free cosmological parameter is $\Omega_{M,0}$, as $\Omega_{\Lambda,0} = 1-\Omega_{M,0}-\Omega_{r,0}$ under the flatness condition.

At the same time, when considering the non-flat curvature extension of the standard $\Lambda \mathrm{CDM}$ model, also the $\Omega_{\Lambda,0}$ is a free parameter, with the present curvature density parameter $ \Omega_{k,0} = 1- \Omega_{M,0} - \Omega_{r,0} - \Omega_{\Lambda,0}$. In this case, we consider the following background evolution
\begin{equation}\label{lcdmnonflat}
E(z) = \frac{H(z)}{H_0} =\Bigg[\Omega_{M,0}\,(1+z)^{3} + \Omega_{r,0}\,(1+z)^{4} + \Omega_{k,0}\,(1+z)^{2} + \Omega_{\Lambda,0}\Bigg]^{\frac{1}{2}}\,
\end{equation}
also including a constraint to rule out the region of the parameters space $(\Omega_{M,0}, \Omega_{\Lambda,0})$ that does not admit an initial singularity, the so-called no Big Bang region. Following \citet{1992ARA&A..30..499C}, we impose
\begin{equation}\label{nobigbang}
\Omega_{\Lambda,0} < 4 \, \Omega_{M,0}\, \mathrm{coss}^{3} \Bigg[\frac{1}{3} \, \mathrm{arccoss} \left( \frac{1- \Omega_{M,0}}{\Omega_{M,0}}\right)\Bigg]
\end{equation}
where $\mathrm{coss}=\mathrm{cos}$ if $\Omega_{M,0}> 1/2$ and $\mathrm{coss}=\mathrm{cosh}$ if $\Omega_{M,0}< 1/2$.

Both for the flat and non-flat cases, and also for all the models presented in the following sections, $k$, $k_{1}$ (only if BAO are fitted), $\gamma$, $\beta$, and $\delta$ are free parameters of the fit, in addition to the specific cosmological parameters. The conservative choice of marginalising also over the slope $\gamma$ is explained in detail in \citet{2020A&A...642A.150L}.

\subsection{Extensions to the $\Lambda \mathrm{CDM}$ model}

Extensions of the standard model also involve exploring forms of DE other than the simple cosmological constant \citep[see e.g.][]{RoccoReview}. In general, we can use a parameterisation of $w(z)$ which can be constant in time or redshift-dependent.
In what follows, we propose to analyse three models well-known in the literature, namely the $w$CDM, CPL and JBP, and two models constructed from specific physical and theoretical assumptions, such as 
i) the DE component must be dominant only at very low redshifts and negligible at early epochs, due to the radiation- and matter-dominated regimes that have preceded the recent DE domination along the Universe evolution; this implies a well-bounded behaviour of DE; ii) $w(z)$ must be negative to give rise to negative pressure and repulsive force able to counter the gravitational attractive force; in particular, according to the second Friedmann's equation, we need $ w(z)<-1/3$ at low $z$ to explain the present accelerated expansion of the Universe \citep{riess1998,perlmutter1999} as a DE domination effect. Specific constraints on the future behaviour of DE ($z \rightarrow (-1)^{+}$) are not required as, generically, cosmological models can show convergence or singularity (e.g. CPL parameterisation) in this limit. These requirements draw inspiration for the ``exponential'' and BA models introduced in this section and already presented in \citet{2019PhRvD..99d3543Y} and \citet{2008PhLB..666..415B}, respectively.
\footnote{All the models studied in this work do not verify the requirements a priori, but only for specific ranges of values of their free parameters.}.

In full generality, for all the DE extensions in this work, we use the following expression for $\displaystyle E(z) = {H(z)}/{H_{0}}$, with a generic $w(z)$:
\begin{equation} \label{E(z)}
\begin{split}
E(z) =& \Bigg[ \Omega_{M,0}\,(1+z)^{3} + \Omega_{r,0}\,(1+z)^{4} + \Omega_{k,0}\,(1+z)^{2} +\\& + \Omega_{\Lambda,0}\,\mathrm{exp}\left(3 \int_{0}^{z} dz'\frac{1+w(z')}{1+z'}\right)\Bigg]^{\frac{1}{2}}.
\end{split}
\end{equation}
Accordingly, the model-dependent luminosity distance we use to derive the predicted distance moduli is obtained from\footnote{This formula is obviously valid also in the flat and non-flat $\Lambda$CDM cases described before.}
\begin{equation} 
\label{Dl(z)}
\displaystyle
 D_\mathrm{{L}}(z)= \begin{cases}
 \frac{c}{H_{0}} (1+z) \frac{\sinh\left[\sqrt{\Omega_{k,0}} \int_{0}^{z}\frac{dz'}{E(z')}\right]}{\sqrt{\Omega_{k,0}}} &\Omega_{k,0}>0,\\
 \frac{c}{H_{0}} (1+z)\int_{0}^{z}\frac{dz'}{E(z')}  &\Omega_{k,0}=0,\\
 \frac{c}{H_{0}} (1+z) \frac{\sin\left[\sqrt{-\Omega_{k,0}} \int_{0}^{z}\frac{dz'}{E(z')}\right]}{\sqrt{-\Omega_{k,0}}} &\Omega_{k,0}<0.\\
\end{cases}
\end{equation}
For all the DE extensions studied, except for the $w$CDM one, we assume a flat Universe fixing $\Omega_{k,0} = 0$ and, as a consequence, $\Omega_{\Lambda,0} = 1-\Omega_{M,0}-\Omega_{r,0}$. This assumption is consistent with the recent results in \citet{eboss2021} and \citet{2021JCAP...11..060G}, where non-flat universes with varying DE EoS are consistent with zero curvature.
In the following, we describe the extended models studied in this work specialising Eq.\eqref{E(z)} for each one.

\subsubsection{Flat and non-flat $w$CDM model}
Apart from considering a non-zero curvature, the simplest and most natural extension of the $\Lambda \mathrm{CDM}$ model is the one in which $w$ is still constant in time but it can assume values different from $w=-1$ and allowed by the constraint $w<-1/3$. This is the so-called $w$CDM model, in which we have
\begin{equation}\label{wcdm}
E(z) = \Bigg[\Omega_{M,0}\,(1+z)^{3} + \Omega_{k,0}\,(1+z)^{2}+ \Omega_{r,0}\,(1+z)^{4} +
\Omega_{\Lambda,0}\, (1+z)^{3(1+w)}\Bigg]^{\frac{1}{2}}
\end{equation}
where $\Omega_{M,0}$ and $w$ are the free parameters under the flatness assumption, together with the additional $\Omega_{\Lambda,0}$ in the non-flat case. The regime $w>-1$ is referred to as ``quintessence'', while the one with $w<-1$ as ``phantom''. The phantom DE scenario predicts a final ``Big Rip'' for the Universe in which all the matter is ripped apart by the accelerated expansion. This DE classification applies also to the models discussed in the following sections, in which $w$ evolves with redshift.

\subsubsection{CPL model}
The most widely used parameterisation for a DE EoS that evolves with time is the CPL \citep{CHEVALLIER_2001,2003PhRvL..90i1301L}, in which $ w(z) = w_{0} + w_{a}\,z/(1+z)$. This is a smooth function in the redshift range $[0,+\infty)$ that starts from $w=w_{0}+w_{a}$ at $z = +\infty$ and reaches $w_{0}$ at $z=0$, but it diverges in the future as $w(z) = \pm \infty$ for $z \rightarrow (-1)^{+}$. Substituting in Eq.\eqref{E(z)}, we get
\begin{small}
\begin{equation}\label{cpl}
E(z) = \Bigg[\Omega_{M,0}\,(1+z)^{3} + \Omega_{r,0}\,(1+z)^{4} +
\Omega_{\Lambda,0}\, (1+z)^{3(1+w_{0}+w_{a})} \mathrm{exp}\left(\frac{-3 w_{a}z}{1+z}\right)\Bigg]^{\frac{1}{2}}
\end{equation}
\end{small}
where $\Omega_{M,0}$, $w_{0}$ and $w_{a}$ are the cosmological free parameters.

\subsubsection{JBP model}
To explore a parameterisation with different dependence on scale, we consider here the JBP model \citep{2005MNRAS.356L..11J} where $w(z) = w_{0} + w_{a} \, z/(1+z)^{2}$ and $w(z)$ has the same value ($w_{0}$) at present epoch ($z=0$) and remote past ($z = +\infty$) with a rapid variation at low-$z$. This model presents the same divergence in the future as the CPL described above. From Eq.\eqref{E(z)} we derive
\begin{small}
\begin{equation}\label{jbp}
E(z) = \Bigg[\Omega_{M,0}\,(1+z)^{3} + \Omega_{r,0}\,(1+z)^{4} +
\Omega_{\Lambda,0}\, (1+z)^{3(1+w_{0})} \mathrm{exp}\left(\frac{3 w_{a}z^{2}}{2(1+z)^{2}}\right)\Bigg]^{\frac{1}{2}}.
\end{equation}
\end{small}
As for CPL model, $\Omega_{M,0}$, $w_{0}$ and $w_{a}$ are the free parameters.

\subsubsection{Exponential model}
This parameterisation, already introduced in \citealt{2019PhRvD..99d3543Y}, assumes
$w(z) = \frac{w_{0}}{1+z}e^{\frac{z}{1+z}}$ and satisfies (for specific values of $w_0$) the physical and theoretical conditions described before. By construction, $w(z=0) = w_{0}$ and $w(z)=0$ for both $z=+\infty$ and $z \rightarrow (-1)^{+}$. The corresponding $E(z)$ is
\begin{small}
\begin{equation}\label{expmodel}
\displaystyle E(z) = \Bigg\{\Omega_{M,0}\,(1+z)^{3} + \Omega_{r,0}\,(1+z)^{4} +
\Omega_{\Lambda,0}\, (1+z)^{3} \mathrm{exp} \Bigg[3 w_{0}\left(e^{\frac{z}{1+z}}-1\right)\Bigg]\Bigg\}^{\frac{1}{2}}.
\end{equation}
\end{small}In this case, we only have $\Omega_{M,0}$ and $w_{0}$ as cosmological free parameters.

\subsubsection{BA model}
Finally, we propose the second model that satisfies the already introduced specific physical and theoretical conditions, defined by the EoS
$w(z) = w_{0} + w_{a} \frac{z\,(1+z)}{1+z^{2}}$. In the asymptotic limits of this model we have $w(z)=w_{0}+w_{a}$ for $z=+\infty$ and $w(z)=w_{0}$ in the future. This is a completely smooth function, without any singularity, in which $\Omega_{M,0}$, $w_{0}$, and $w_{a}$ are the free parameters. In this case, from Eq.\eqref{E(z)} we get
\begin{small}
\begin{equation}\label{rationalmodel}
E(z) = \Bigg[\Omega_{M,0}\,(1+z)^{3} + \Omega_{r,0}\,(1+z)^{4} +
\Omega_{\Lambda,0}\, (1+z)^{3(1+w_{0})} (1+z^{2})^{\frac{3}{2}w_{a}}\Bigg]^{\frac{1}{2}}.
\end{equation}
\end{small}

\begin{figure}
    \resizebox{\hsize}{!}{\includegraphics{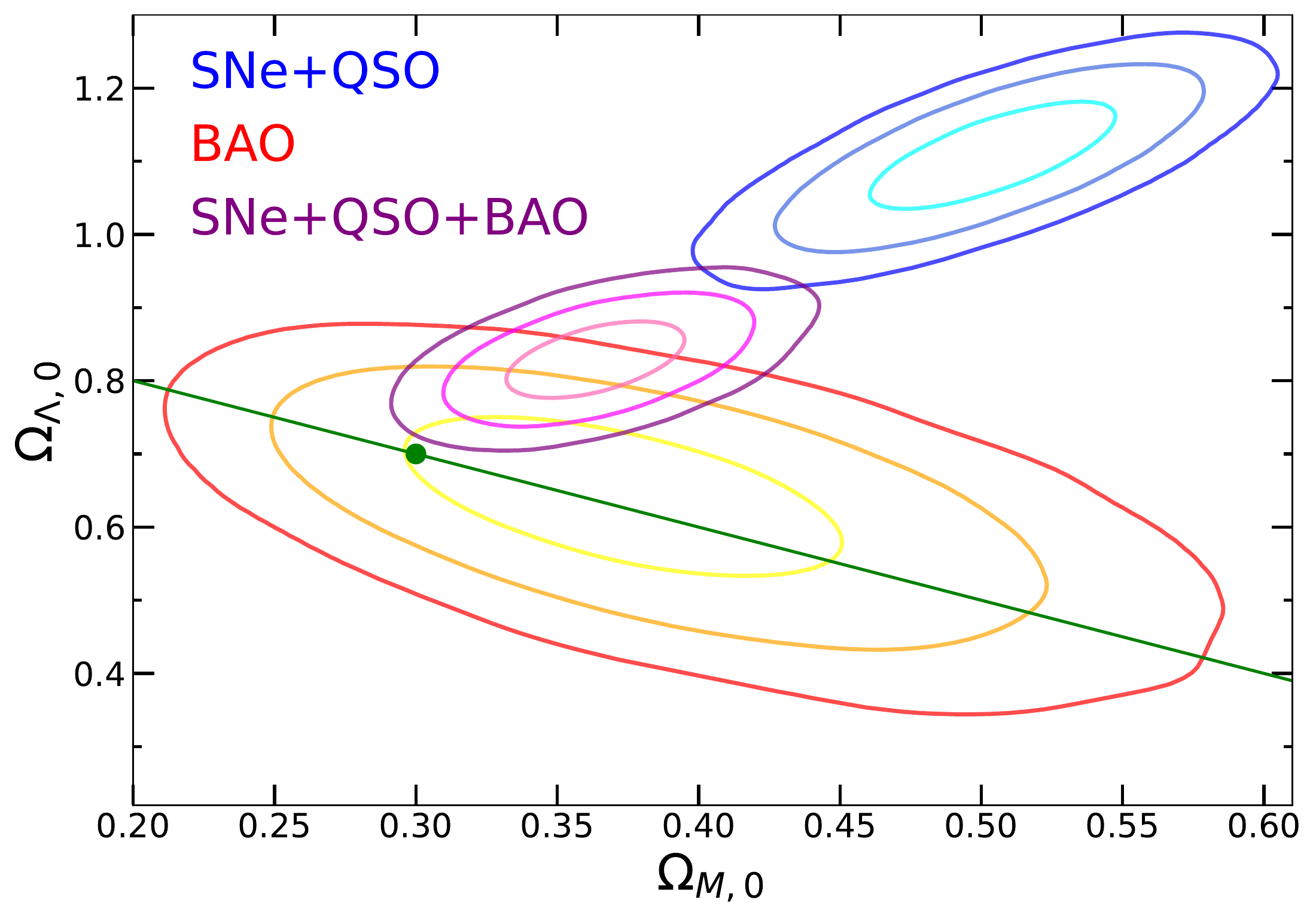}}
    \caption{ Bi-dimensional contour plot ($\Omega_{M,0}$, $\Omega_{\Lambda,0}$) with 1-3$\sigma$ confidence levels obtained from each data set in the legend for the non-flat $\Lambda \mathrm{CDM}$ model analysis. The green line is the place of points corresponding to a flat model and the green point on it is located at $\Omega_{M,0}=0.3$ and $\Omega_{\Lambda,0}=0.7$, close to the values expected in a flat $\Lambda \mathrm{CDM}$ model. The confidence levels for the SNe+QSO+BAO data set are shown here just for sake of clarity as the two probes show strong incompatibility and their combined analysis is not meaningful (see text). 
    }
    \label{nonflatlcdmqso+sne+bao}
\end{figure}

\section{Cosmological results}
\label{Cosmological results}


In this section, we combine the data introduced above by considering the data sets of SNe+QSO, BAO alone, and SNe+QSO+BAO to analyse the models with curvature and DE extension just described, and discuss the compatibility of the data or their tension.
As already mentioned, we use the most recent QSO sample carefully and specifically selected for cosmological purposes \citep{2020A&A...642A.150L}. It covers a wide redshift range up to $z \sim 7.5$, allowing us to extend the Hubble diagram in a high-$z$ region not explored by SNe (that reach a maximum redshift of 2.26). This is a crucial region to distinguish among different cosmological models, that are instead degenerate at low redshifts. For these reasons, QSOs are expected to be responsible for a shift in the free parameter spaces of the models (due to information at higher redshifts) and improved constraints on the parameters (due to the update of the sample) compared to previous works \citep[e.g.][]{rl15, rl19}. 
These expectations have already been confirmed in \citet{refId0}, where the same sample of QSOs (without the inclusion of BAO) is used for a cosmographic test of the flat $\Lambda \mathrm{CDM}$ model, with an improved estimation of the cosmographic free parameters, compared to the use of less updated QSO samples. 
\begin{figure}
    \resizebox{\hsize}{!}{\includegraphics{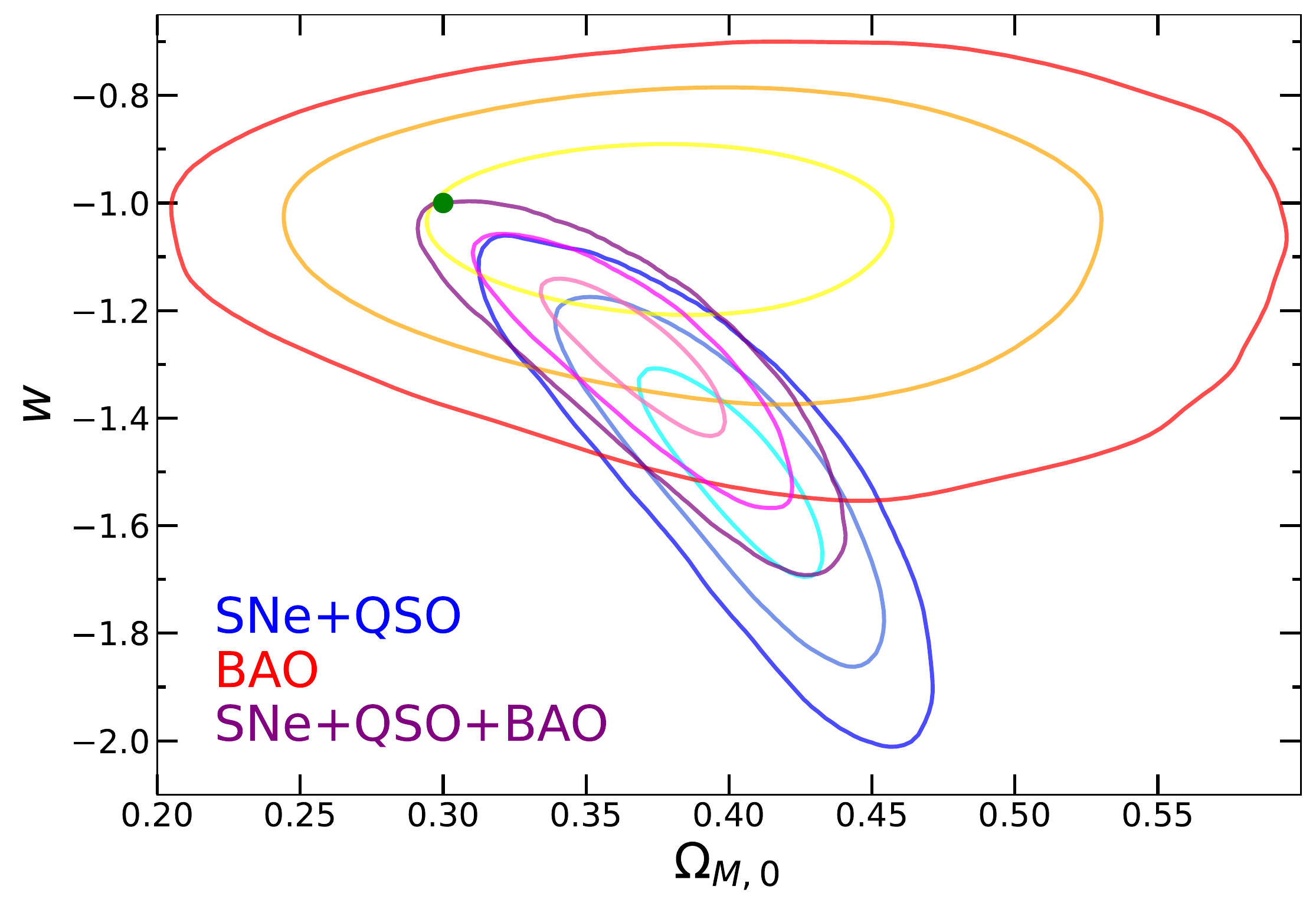}}
    \caption{Bi-dimensional contour plot ($\Omega_{M,0}$, $w$) with 1-3$\sigma$ confidence levels obtained from each data set in the legend for the flat $w \mathrm{CDM}$ model analysis. The green point is located at $\Omega_{M,0}=0.3$ and $w=-1$, corresponding to the cosmological constant case.}
    \label{wcdmqso+sne+bao}
\end{figure}

It is worth mentioning some of the assumptions used in the cosmological analyses before moving on. First of all, we use
flat uniform priors on the free parameters: $ 0 \leq \Omega_{M,0} \leq 1$, $-\infty < w \leq 0$ and, when fitting also QSOs, $0\leq \gamma \leq 1$ and $-10 < \mathrm{ln} \, \delta < 1$. 
However, there are some exceptions to this rule concerning when CPL and BA models are analysed with SNe+QSO sample. In these cases, we require $ 0.32 \leq \Omega_{M,0} \leq 1$ and $ 0.34 \leq \Omega_{M,0} \leq 1$, respectively, and the reasons are discussed in Sects. \ref{resultscpl}, \ref{resultsrational} and Appendix \ref{appendixcpl}. 
Then, in the non-flat $\Lambda \mathrm{CDM}$ model we also always include the no Big Bang constraint according to Eq. \eqref{nobigbang}, while fitting the non-flat $w$CDM on the BAO sample we restrict $w$ such that $-5<w<0$ to bound the lower limit. 

\begin{figure}
     \resizebox{\hsize}{!}{\includegraphics{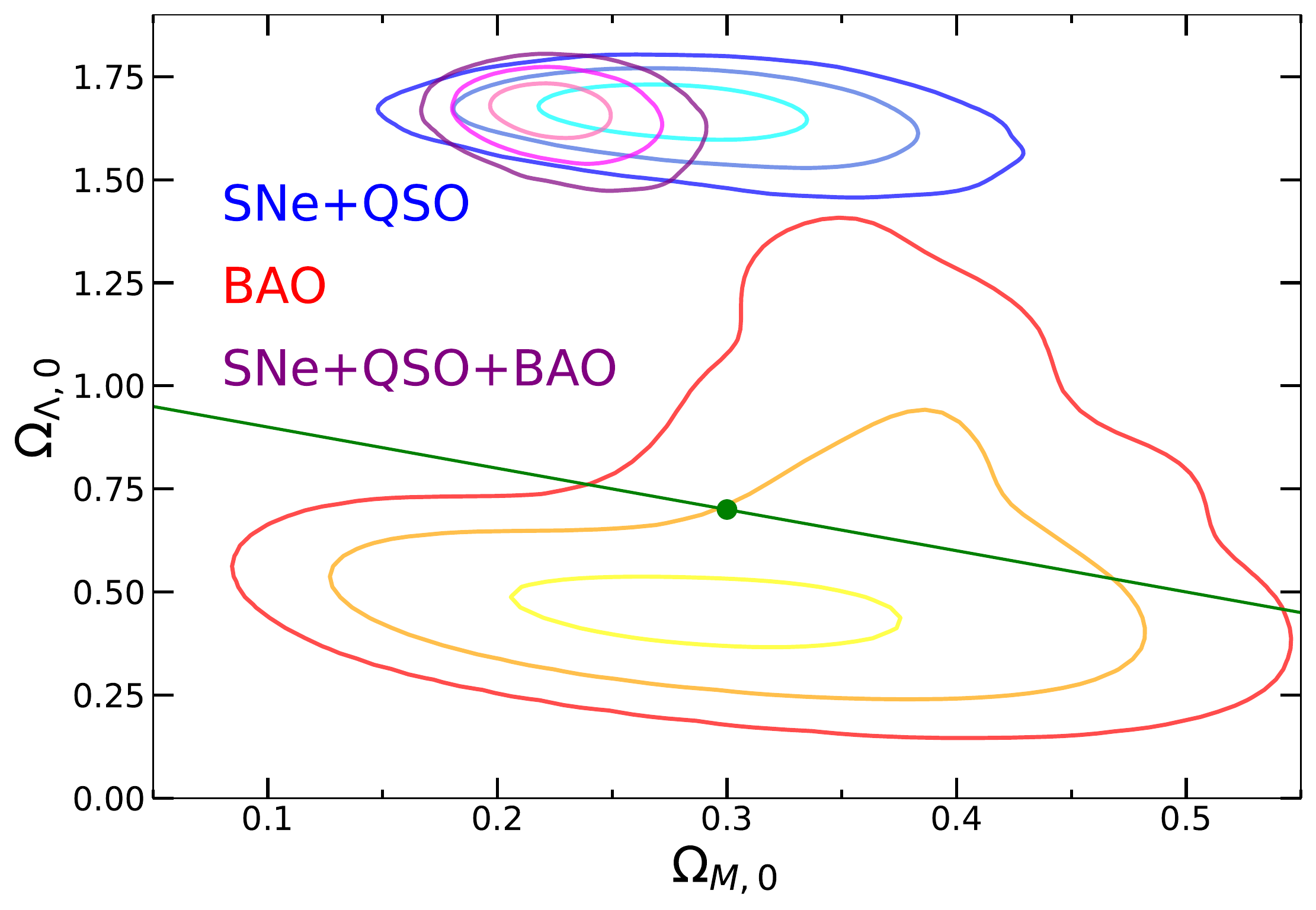}}
    \caption{Bi-dimensional contour plot ($\Omega_{M,0}$, $\Omega_{\Lambda,0}$) with 1-3$\sigma$ confidence levels obtained from each data set in the legend for the non-flat $w \mathrm{CDM}$ model analysis. The green line is the place of points corresponding to a flat model and the green point on it is located at $\Omega_{M,0}=0.3$ and $\Omega_{\Lambda,0}=0.7$, close to the values expected in a flat $\Lambda \mathrm{CDM}$ model. The confidence levels for the SNe+QSO+BAO data set are shown here just for sake of clarity as the two probes show strong incompatibility and their combined analysis is not meaningful (see text).} 
    \label{nonflatwcdmqso+sne+bao}
\end{figure}

Our results are summarised in Table \ref{tablefits}, where we present the mean values with 1$\sigma$ error.
These are obtained applying to the QSO sample the sigma-clipping procedure\footnote{This method is commonly used in presence of an intrinsic dispersion to remove possible outliers in the data set and guarantee a better estimation of the free parameters.} relative to the specific model studied; more precisely, we iteratively discard every QSO source with a discrepancy greater than 3$\sigma$ from the best-fit model.
We also show in the figures of this section the behaviours of $\Omega_{\Lambda}$ and $\Omega_M$ in non-flat curvature assumption for the $\Lambda$CDM and $w$CDM models, as well as the dependence of EoS parameters both on $\Omega_M$ and other DE parameter (see Figs. \ref{nonflatlcdmqso+sne+bao}, \ref{wcdmqso+sne+bao}, \ref{nonflatwcdmqso+sne+bao} \ref{cplqso+sne+bao}, \ref{expqso+sne+bao}, \ref{jbpqso+sne+bao}, and \ref{rationalqso+sne+bao}, respectively).
The confidence levels are always the 1, 2, and 3$\sigma$ intervals, corresponding to 68.3, 95, and 99.7\% confidence level, respectively.

We point out that the $H_0$ value has been fixed in all the models subject to our analysis. This leads to some differences (e.g. the degeneracy direction of the confidence contours shown in Fig. \ref{nonflatlcdmqso+sne+bao} for BAO data in the non-flat $\Lambda$CDM model) in our results with other similar analyses already presented in the literature \citep[e.g.][]{eboss2021,2021JCAP...11..060G}. Also, our fitting procedure takes on a simpler treatment of the cosmological model than CosmoMC \citep{Lewis:2002ah} and Montepython \citep{Audren:2012wb} algorithms (which are among the most commonly used codes). Therefore, care should be taken in comparing our findings with previous works in the literature that make use of the cosmological packages listed above, especially when employed with different data sets and calibrations of the probes. The implementation of the presented analysis in CosmoMC and Montepython is being studied and will be presented in a forthcoming paper.

Our analysis shows that the constraints on the cosmological parameters obtained from the SNe+QSO+BAO sample are not always a simple average of the ones derived from SNe+QSO and BAO separately (see Table~\ref{tablefits}).
In other words, the 2D contours of the combined sample do not necessarily represent the geometric intersection of the individual projections. This effect is due to the relative orientation of the contours in the multi-dimensional space defined by the free model parameters. 
The overall consistency for the SNe+QSO and BAO data sets thus requires the individual constraints to be compatible in the full multi-dimensional space, as the mere accordance of pairs of cosmological parameters could be misleading \citep[see][]{2021JCAP...11..060G}. 
The combination of these two data sets is then likely possible (and meaningful) only if the derived constraints are consistent, which means that the separate confidence levels overlap (within approximately 2$\sigma$) both in the marginalised posterior distributions of each of the shared free parameters and in the 2D projections of each pair of them. 
By comparing both the 1D and 2D information, we find that only the non-flat models show a significant discrepancy between the two probes, as explained in the following Sects. \ref{sectionnonflatresults} and \ref{sectionnonflatwresults}.
Simulations of posterior distributions in a 3D parameter space may help in understanding possible issues, yet any analysis based on projections from a multi-dimensional space with more than three dimensions is surely non-trivial.
Below we summarise the key results of our analysis.
\begin{table*}
\caption{Best-fit values and 1$\sigma$ uncertainties for the cosmological free parameters in each model and data set. The symbol $^{\star}$ identifies the data sets for which the analysis is not statistically justified due to the tension between the probes (even if results are reported for sake of clarity).}
\label{tablefits}
\centering 
\setlength{\tabcolsep}{6pt}
\renewcommand{\arraystretch}{1.5}
\begin{tabular}{ c c c c c c} 
\hline
Model & Data set & $\Omega_{M,0}$ & $\Omega_{\Lambda,0}$ & $w_{0}$ & $w_{a}$\\
\hline
Flat $\Lambda \mathrm{CDM}$ & SNe+QSO & $0.295 ^{+0.013}_{-0.012}$\\
& BAO & $0.373 ^{+0.056}_{-0.048}$ & & &\\
& SNe+QSO+BAO & $0.300 \pm 0.012$ & & & \\
Non-flat $\Lambda \mathrm{CDM}$ & SNe+QSO & $0.504 \pm 0.029$ & $1.107 ^{+0.051}_{-0.052}$ & & \\
& BAO & $0.376 ^{+0.057}_{-0.049}$ & $0.638 ^{+0.071} _{-0.079}$ & & \\
& SNe+QSO+BAO $^{\star}$ & $0.364 ^{+0.022}_{-0.021}$ & $0.829 \pm 0.035$ & & \\
Flat $w$CDM & SNe+QSO & $0.403^{+0.022}_{-0.024}$ & & $-1.494 ^{+0.132}_{-0.143}$ & \\
& BAO & $0.381 ^{+0.057}_{-0.050}$ & & $-1.049 ^{+0.098} _{-0.116}$ & \\
& SNe+QSO+BAO & $0.369 ^{+0.022}_{-0.023}$ & & $-1.283^{+0.094}_{-0.108}$ &\\
Non-flat $w$CDM & SNe+QSO & $0.280 ^{+0.041}_{-0.037}$ & $1.662^{+0.041}_{-0.048}$ & $-0.667^{+0.024}_{-0.027}$ & \\
& BAO & $0.301^{+0.080}_{-0.072}$ & $0.463^{+0.072}_{-0.058}$ & $-2.850^{+1.459}_{-1.441}$ &\\
& SNe+QSO+BAO $^{\star}$ & $0.224^{+0.018}_{-0.017}$ & $1.667^{+0.040}_{-0.047}$ & $-0.626^{+0.012}_{-0.013}$ &\\
CPL & SNe+QSO & $0.447 ^{+0.023}_{-0.027}$ & & $-1.267 ^{+0.196}_{-0.191}$ & $-3.771^{+2.113}_{-2.496}$\\
& BAO & $0.420^{+0.073}_{-0.070}$ & & $-0.821^{+0.469}_{-0.349}$ & $-1.269^{+1.835}_{-2.608}$\\
& SNe+QSO+BAO & $0.354^{+0.032}_{-0.030}$ & & $-1.323^{+0.103}_{-0.112}$ & $0.745^{+0.483}_{-0.974}$\\
JBP & SNe+QSO & $0.441^{+0.025}_{-0.028}$ & & $-1.250^{+0.223}_{-0.209}$ & $-4.282^{+2.680}_{-3.283}$\\
& BAO & $0.384^{+0.103}_{-0.098}$ & & $-1.091^{+0.923}_{-0.727}$ & $0.235^{+4.922}_{-6.612}$\\
& SNe+QSO+BAO & $0.354^{+0.032}_{-0.030}$ & & $-1.371\pm 0.141$ & $1.127^{+1.293}_{-1.547}$\\
Exponential & SNe+QSO & $0.395^{+0.023}_{-0.026}$ & & $-1.481^{+0.141}_{-0.147}$ & \\
& BAO & $0.371^{+0.058}_{-0.051}$ & & $-1.067 ^{+0.102}_{-0.119}$ & \\
& SNe+QSO+BAO & $0.359 ^{+0.023}_{-0.024}$ & & $-1.271^{+0.092}_{-0.107} $& \\
BA & SNe+QSO & $0.452^{+0.022}_{-0.025}$ & & $-1.316^{+0.172}_{-0.168}$ & $-2.654^{+1.329}_{-1.626}$\\
& BAO & $0.410^{+0.086}_{-0.081}$ & & $-0.930^{+0.464}_{-0.333}$ & $-0.423^{+1.064}_{-1.671}$ \\
& SNe+QSO+BAO & $0.307^{+0.044}_{-0.055}$ & & $-1.303^{+0.115}_{-0.106}$ & $1.010^{+0.152}_{-0.466}$\\
 \hline
\end{tabular}
\end{table*}
\subsection{Constraints on $\Lambda \mathrm{CDM}$ model}
\label{sectionflatlcdm}
The only cosmological free parameter of this model is $\Omega_{M,0}$. The best-fit values obtained with SNe+QSO and BAO are consistent within 2$\sigma$ (see Table \ref{tablefits}) and the combination of the two data sets gives $\Omega_{M,0} = 0.300 \pm 0.012$. This result completely agrees with the latest cosmological evidence \citep[e.g.][]{Hinshaw_2013,planck2018,scolnic2018}.

\subsection{Constraints on non-flat $\Lambda \mathrm{CDM}$ model}
\label{sectionnonflatresults}

In this model, SNe+QSO and BAO data sets show a strong discrepancy on the bound value of cosmological parameters (see Table \ref{tablefits} and Fig. \ref{nonflatlcdmqso+sne+bao}). Indeed, the red confidence levels from BAO alone in Fig. \ref{nonflatlcdmqso+sne+bao} are completely consistent with the flat constraint $\Omega_{\Lambda,0}+\Omega_{M,0}+\Omega_{r,0} = 1$ represented by the green line, even if with a preference for $\Omega_{M,0} > 0.3$ and $\Omega_{\Lambda,0} < 0.7$, corresponding to a 1$\sigma$ discrepancy from the flat $\Lambda \mathrm{CDM}$ model with $\Omega_{M,0} = 0.3$ and $\Omega_{\Lambda,0} = 0.7$ (green point), the values expected from the latest observational evidence. 
On the other hand, the blue contours from SNe+QSO prefer a closed Universe ($\Omega_{k,0}<0$) with high values for both $\Omega_{M,0}$ and $\Omega_{\Lambda,0}$, in agreement with the results in
\citet{2020NatAs...4..196D,divalentino2021}. As already stated in \citet{2021JCAP...11..060G},
in this case, where BAO and SNe+QSO data do not give compatible results in 2$\sigma$, it is not proper to consider the joint data set SNe+QSO+BAO, due to the tension between the two data sets. Nevertheless, just for sake of clarity, Table \ref{tablefits} reports also the mean values from the combined sample and Fig. \ref{nonflatlcdmqso+sne+bao} shows the corresponding constraints in the bi-dimensional free parameters space (purple contours). The no Big Bang constraint described in Sect. \ref{DE EoS parameterisations} does not influence at all on any of the data sets, as it is limited to the region of the parameters space with high $\Omega_{\Lambda,0}$ and low $\Omega_{M,0}$, completely excluded from all the confidence levels.

\subsection{Constraints on flat $w$CDM model}
\begin{figure*}
    \centering
    \includegraphics[width=6.2cm]{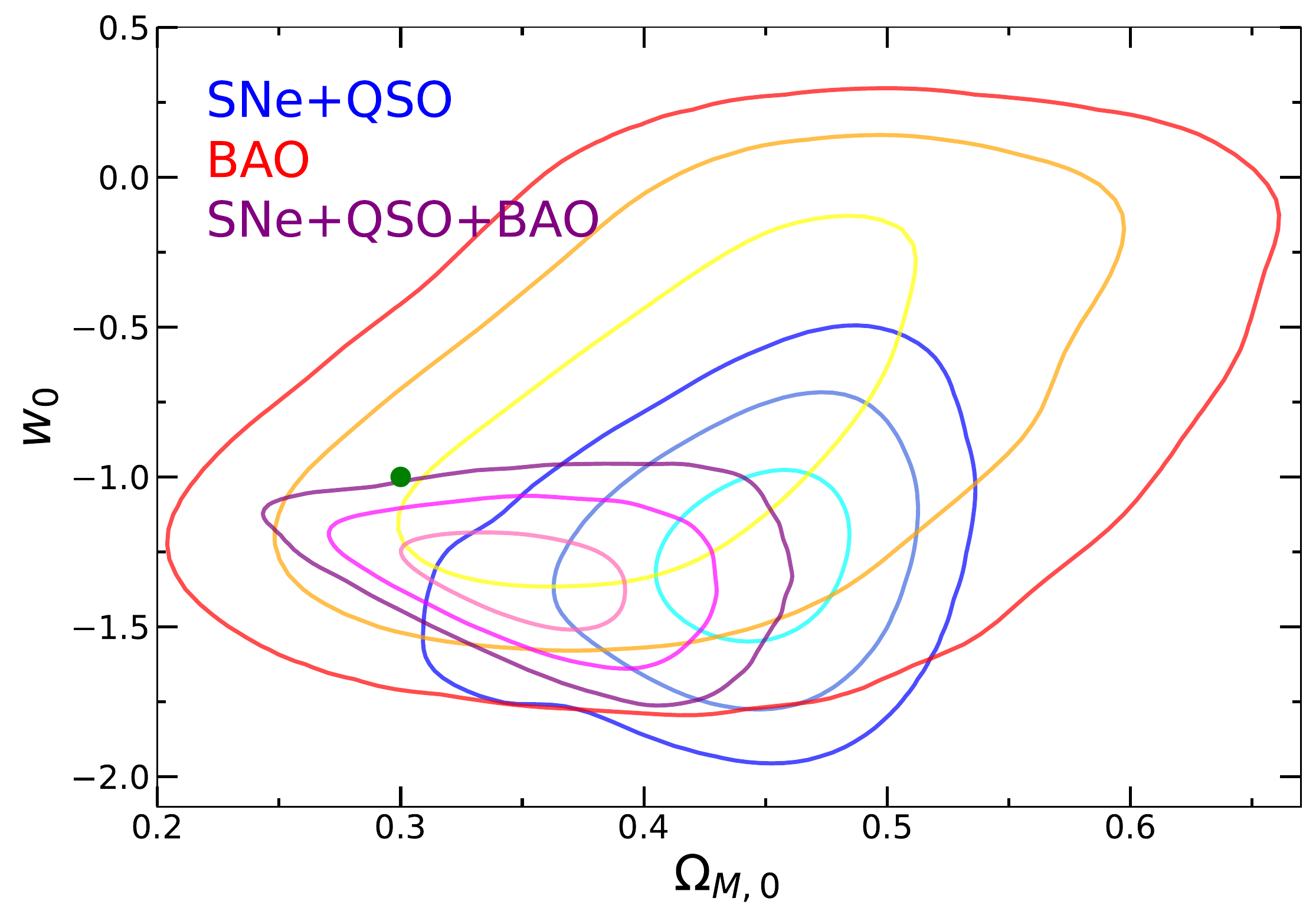}\includegraphics[width=6.2cm]{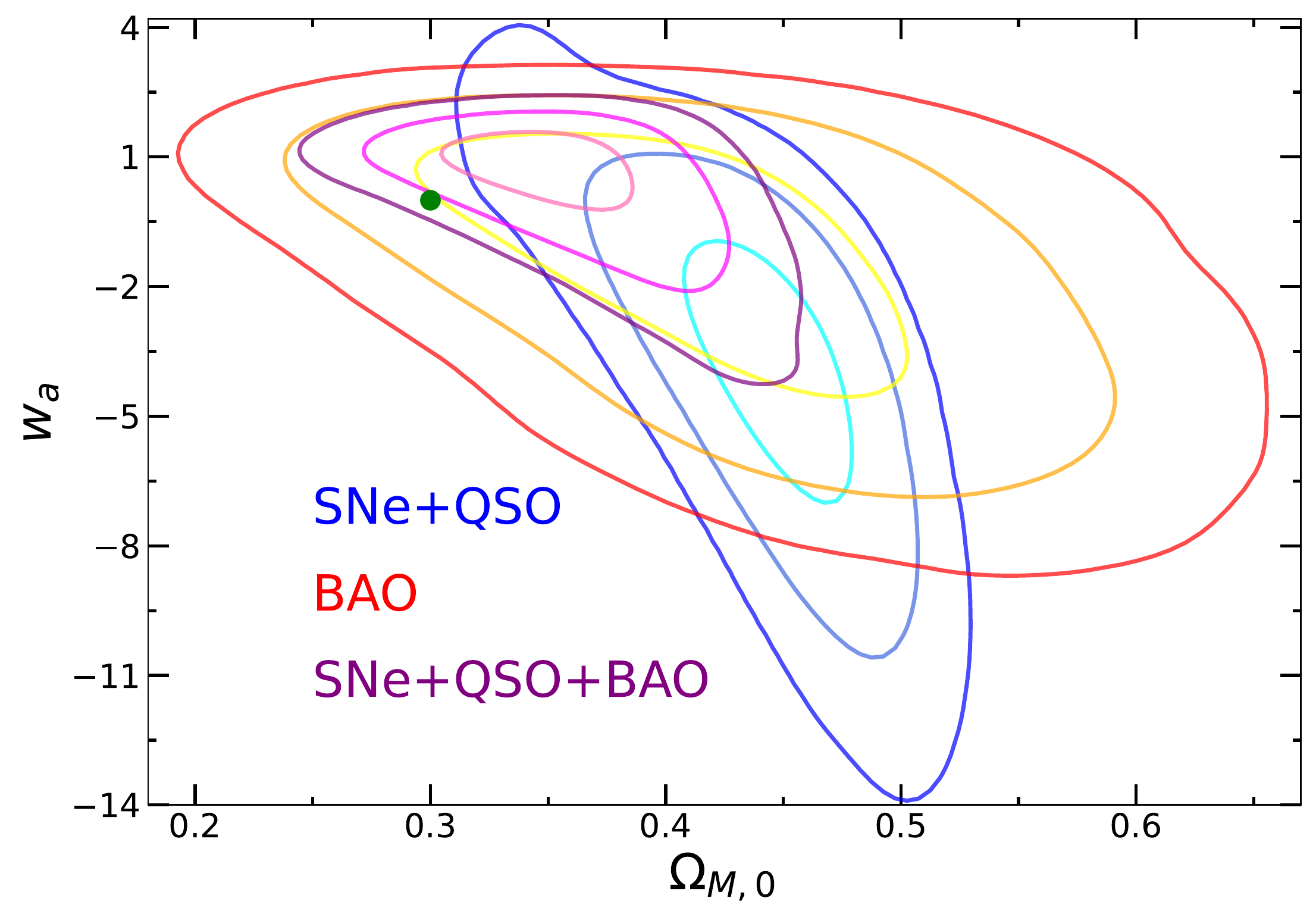}\includegraphics[width=6.2cm]{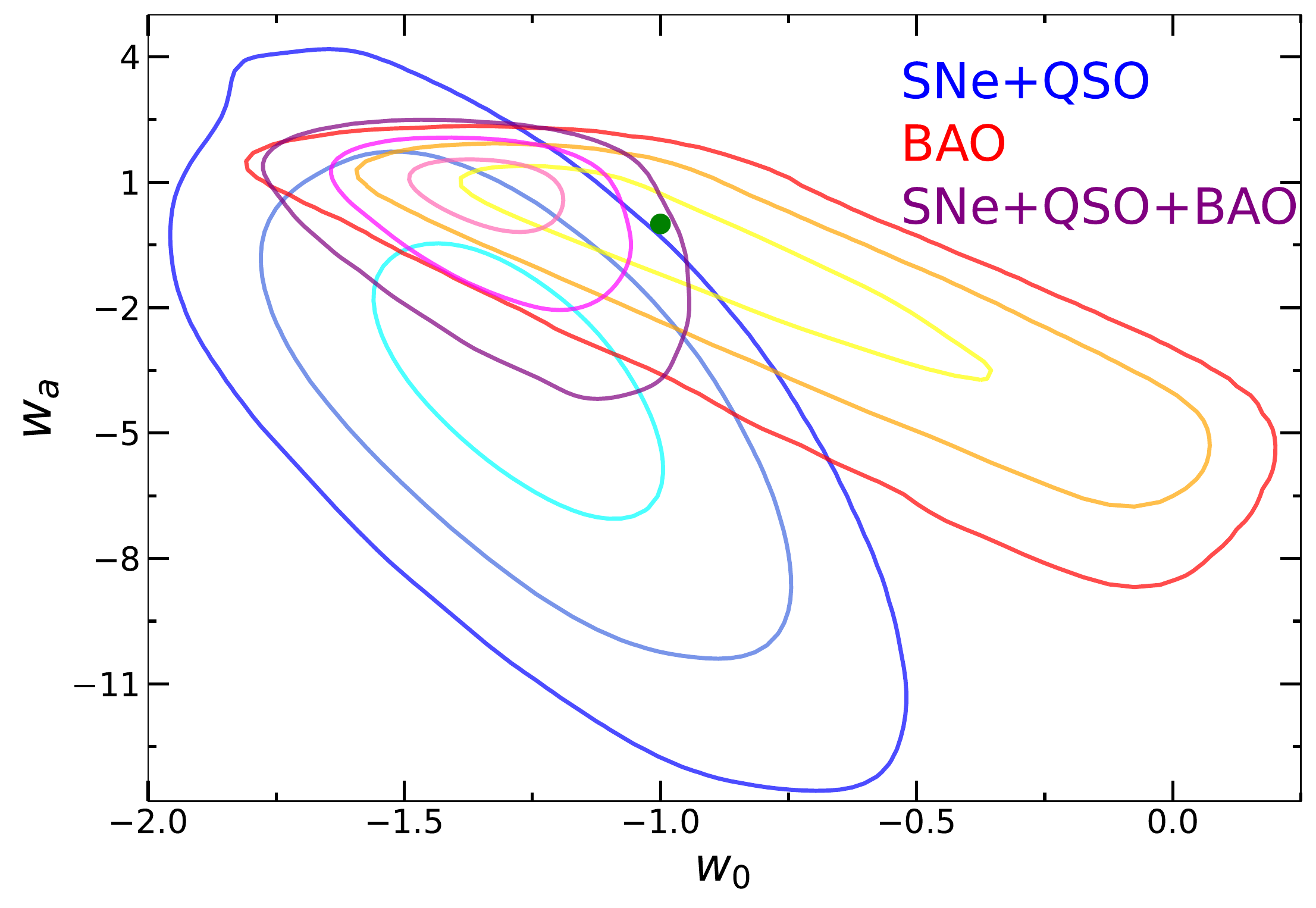}
    \caption{Contour plots for all the pairs of cosmological free parameters in the flat CPL model. The green point corresponds to a flat $\Lambda \mathrm{CDM}$ model with $\Omega_{M,0}=0.3$. The colours-data sets association is reported in the legend.} 
    \label{cplqso+sne+bao}
\end{figure*}
Results for the first and simplest DE extension are reported in Table \ref{tablefits}, while the behaviour of $\Omega_{M,0}$ with $w$ is shown in Fig. \ref{wcdmqso+sne+bao}. The SNe+QSO data set (in blue) shows a tension greater than 3$\sigma$ with the green point located at (0.3,-1), that corresponds to a cosmological constant and $\Omega_{M,0}=0.3$, preferring a higher $\Omega_{M,0}$ and a more negative $w$. The best-fit value $w\sim-1.5$ corresponds indeed to a phantom DE scenario. The BAO data set (in red) instead gives a $w$ distribution consistent with $w=-1$, even if with $\Omega_{M,0}$ always greater than $\Omega_{M,0}=0.3$. The combination of the two data sets, justified by their consistency within 2$\sigma$, retains the same anti-correlation between $\Omega_{M,0}$ and $w$ as SNe+QSO, with a resulting 3$\sigma$ discrepancy from the flat $\Lambda \mathrm{CDM}$ model with $\Omega_{M,0} = 0.3$ that favours a greater value of $\Omega_{M,0}$ and $w<-1$. 

\subsection{Constraints on non-flat $w$CDM model}
\label{sectionnonflatwresults}
In the case where both a constant EoS of DE not fixed to $w=-1$ and a non-flat curvature of the Universe are considered, we obtain that
SNe+QSO and BAO data cannot be joint due to their strong inconsistency. Indeed, Table \ref{tablefits} and Fig. \ref{nonflatwcdmqso+sne+bao} show that the two probes give compatible values of $\Omega_{M,0}$, but completely discrepant values of $\Omega_{\Lambda,0}$ and 2D contour plots ($\Omega_{M,0}$, $\Omega_{\Lambda,0}$) at more than 3$\sigma$ tension. 
Since BAO are not sensitive to the EoS parameter of DE, which means that such a data set cannot constrain $w$, we do not show the behaviour of $w$ with the other parameters;
this is also the reason why we modify the uniform prior on $w$ fitting this model on the BAO data set, requiring $-5<w<0$.
Anyway, we are mostly interested in studying the implication on the spatial curvature of the Universe. About that, as for the non-flat $\Lambda$CDM model, BAO alone are consistent with a flat Universe within 2$\sigma$ with a preference toward $\Omega_{M,0}\sim0.3$ and $\Omega_{\Lambda,0}\sim0.5$, while SNe+QSO prefer a negative curvature (closed Universe) with $\Omega_{M,0}\sim 0.3$ but a high value of $\Omega_{\Lambda,0}$. Despite the inconsistency, as for the non-flat $\Lambda$CDM case, we report in Table \ref{tablefits} and Fig. \ref{nonflatwcdmqso+sne+bao} also the results obtained from the joint sample.

\subsection{Constraints on CPL model}
\label{resultscpl}

This model, together with the flat BA one, requires a more detailed description of the priors. Indeed, as anticipated, we use a very strict prior on $\Omega_{M,0}$ (i.e. $ 0.32 \leq \Omega_{M,0} \leq 1$) when fitting the SNe+QSO data set. The reason for this very specific choice can be clarified by looking at Fig. \ref{2peakcpl}: without this prior, the MCMC exploration of the parameters space is not able to converge properly, as it detects two different families of solutions. Between these two, the one with $\Omega_{M,0} \sim 0$ is obviously non-physical and corresponds to a relative (and not absolute) maximum of the likelihood function explored. Consequently, we remove it with the prior on $\Omega_{M,0}$.

Figure \ref{cplqso+sne+bao} shows the 2D contour plots for all the cosmological free parameters in each of the data sets considered in this work. Indeed, the constraints from SNe+QSO (blue contours) and BAO alone (red contours) are always completely consistent (also due to the large uncertainties) and we can combine them. The green point is the reference of a flat $\Lambda \mathrm{CDM}$ model with $\Omega_{M,0} = 0.3$, as before. We can note that both the data sets prefer $\Omega_{M,0}>0.4$, while we get $w_{0} < -1$ and $w_{a} \sim -4$ from SNe+QSO and $w_{0}$ consistent with -1 and $w_{a} \sim -1.3$ from BAO, at more than 3$\sigma$ and at about 1$\sigma$ from the flat $\Lambda \mathrm{CDM}$ model with $\Omega_{M,0} = 0.3$ for each pair of free parameters, respectively. The joint fit (in purple) ends in tighter constraints with $\Omega_{M,0}$ shifted toward $\Omega_{M,0}=0.3$, $w_{0} <-1$ completely consistent with SNe+QSO and $w_{a}$ slightly positive, with an overall statistical significance of 2-3$\sigma$ relative to the reference flat $\Lambda \mathrm{CDM}$ prediction. The change in sign of $w_{a}$ from negative to positive values (or at least values consistent with $w_{a}=0$) is mainly due to the inclusion of BAO and it agrees with the analysis shown in \citet{2016A&A...594A..13P,planck2018}. Compared to the results of the $w$CDM model, the best-fit values of $\Omega_{M,0}$ and $w$ are completely consistent in each data set, while their correlation is different due to the presence of the additional $w_{a}$ parameter in the CPL model.

\begin{figure*}
    \centering
    \includegraphics[width=6.2cm]{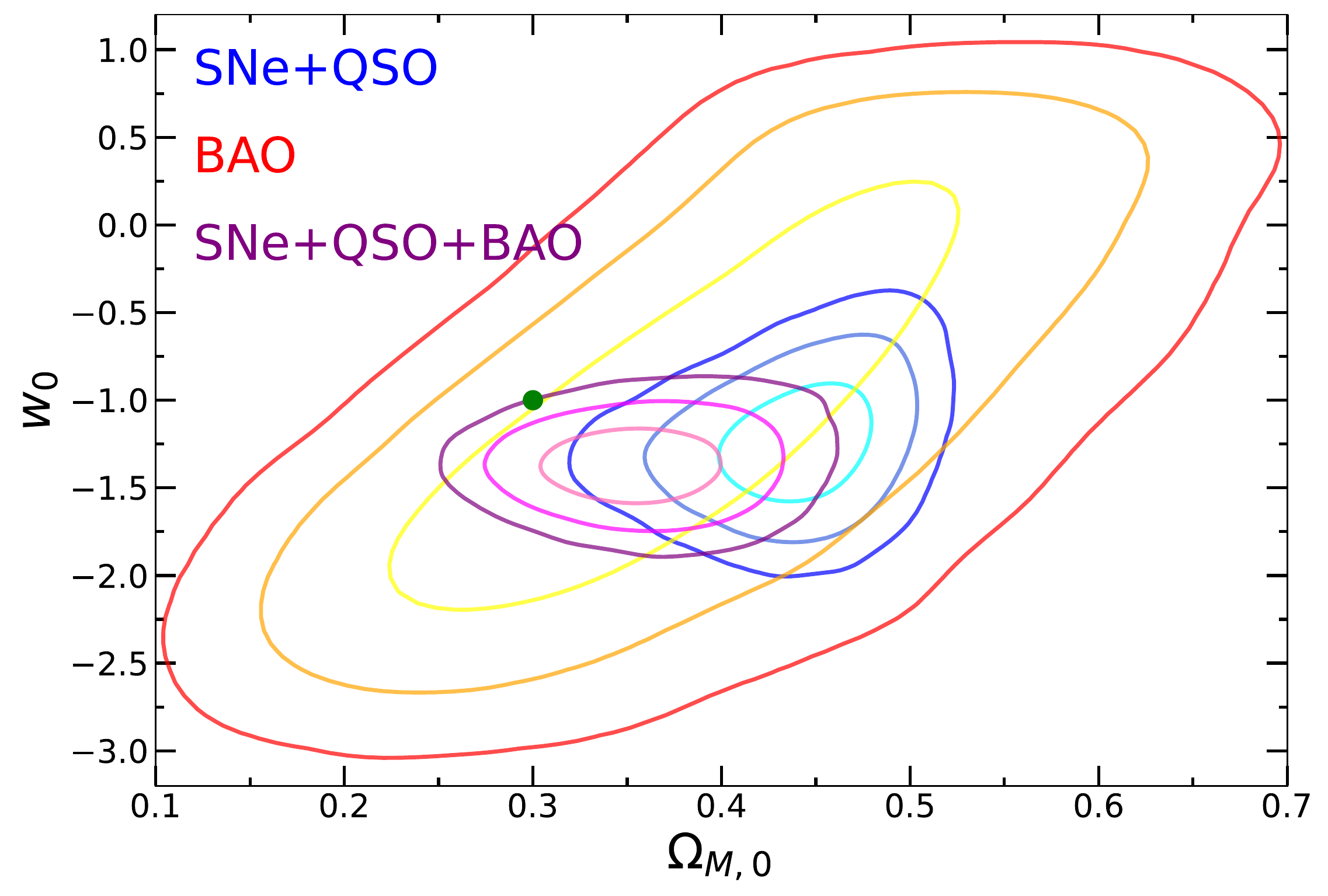}\includegraphics[width=6.2cm]{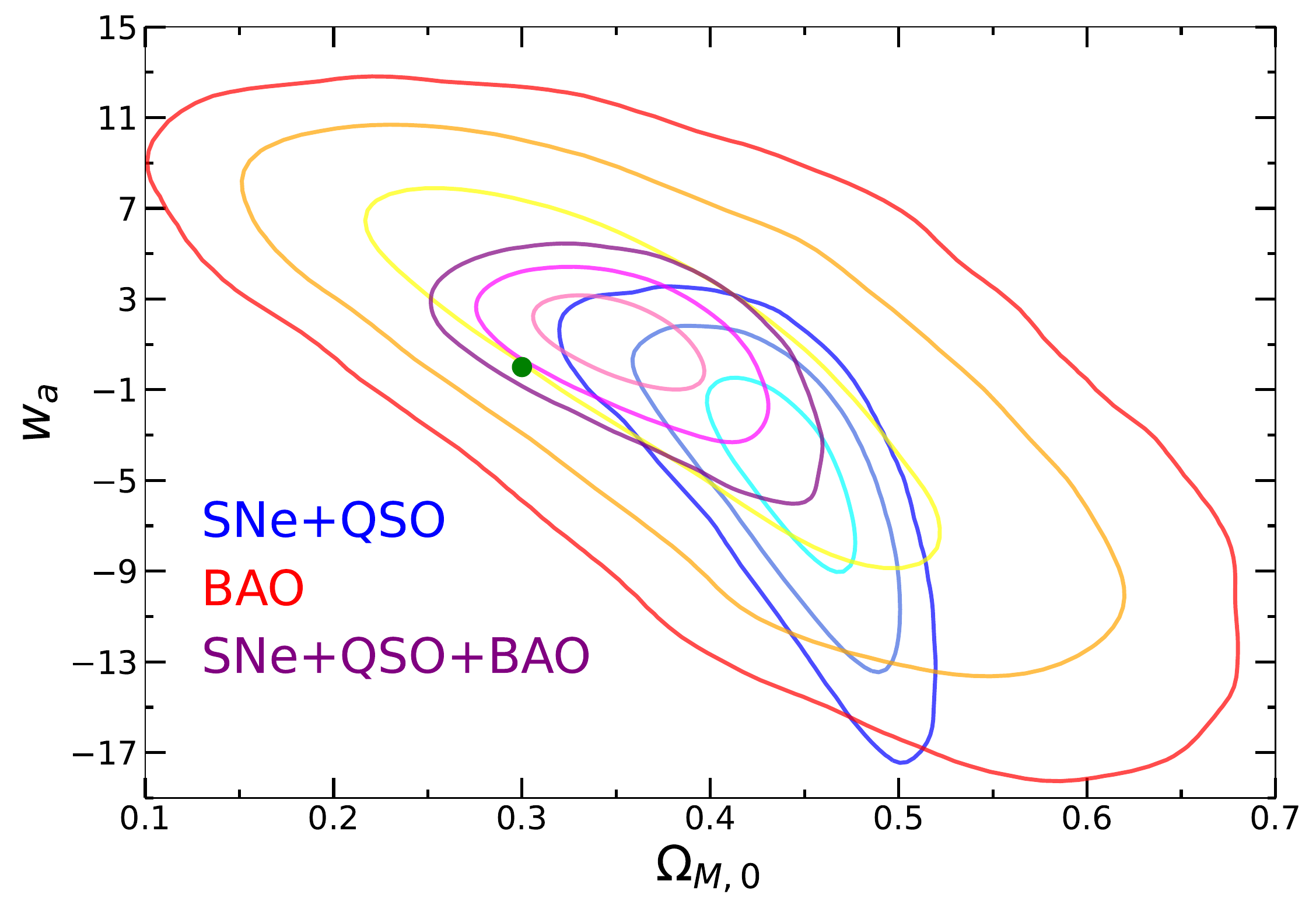}\includegraphics[width=6.2cm]{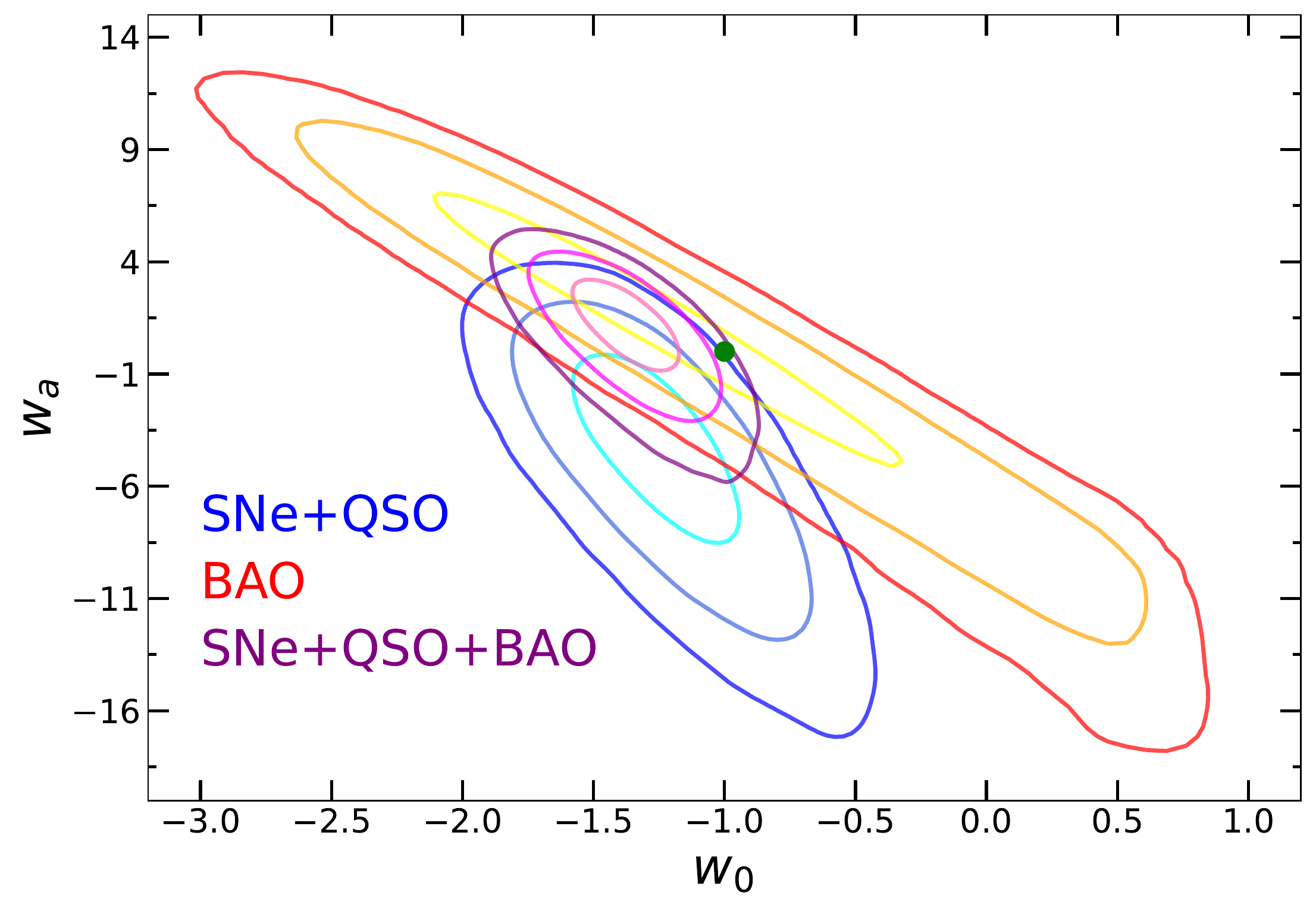}
    \caption{Bi-dimensional contour plots with 1-3$\sigma$ confidence levels obtained from each data set in the legend for the flat JBP model.  The green point corresponds to a flat $\Lambda \mathrm{CDM}$ model with $\Omega_{M,0}=0.3$.} 
    \label{jbpqso+sne+bao}
\end{figure*}

\begin{figure}
    \resizebox{\hsize}{!}{\includegraphics{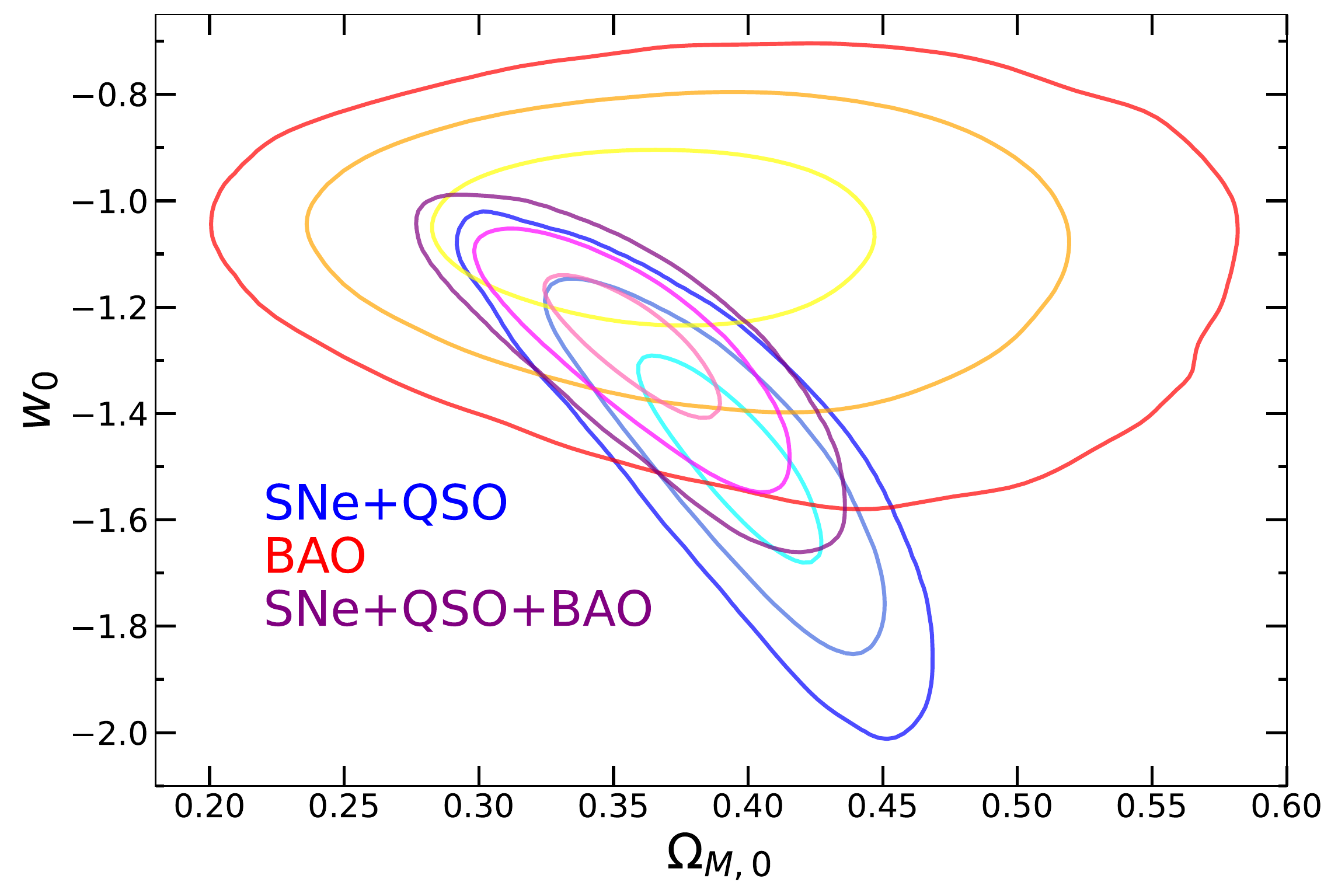}}
    \caption{
    Bi-dimensional contour plot ($\Omega_{M,0}$, $w_{0}$) with 1-3$\sigma$ confidence levels obtained from each data set in the legend for the flat exponential model.}
    \label{expqso+sne+bao}
\end{figure}

\begin{figure*}
    \centering
    \includegraphics[width=6.2cm]{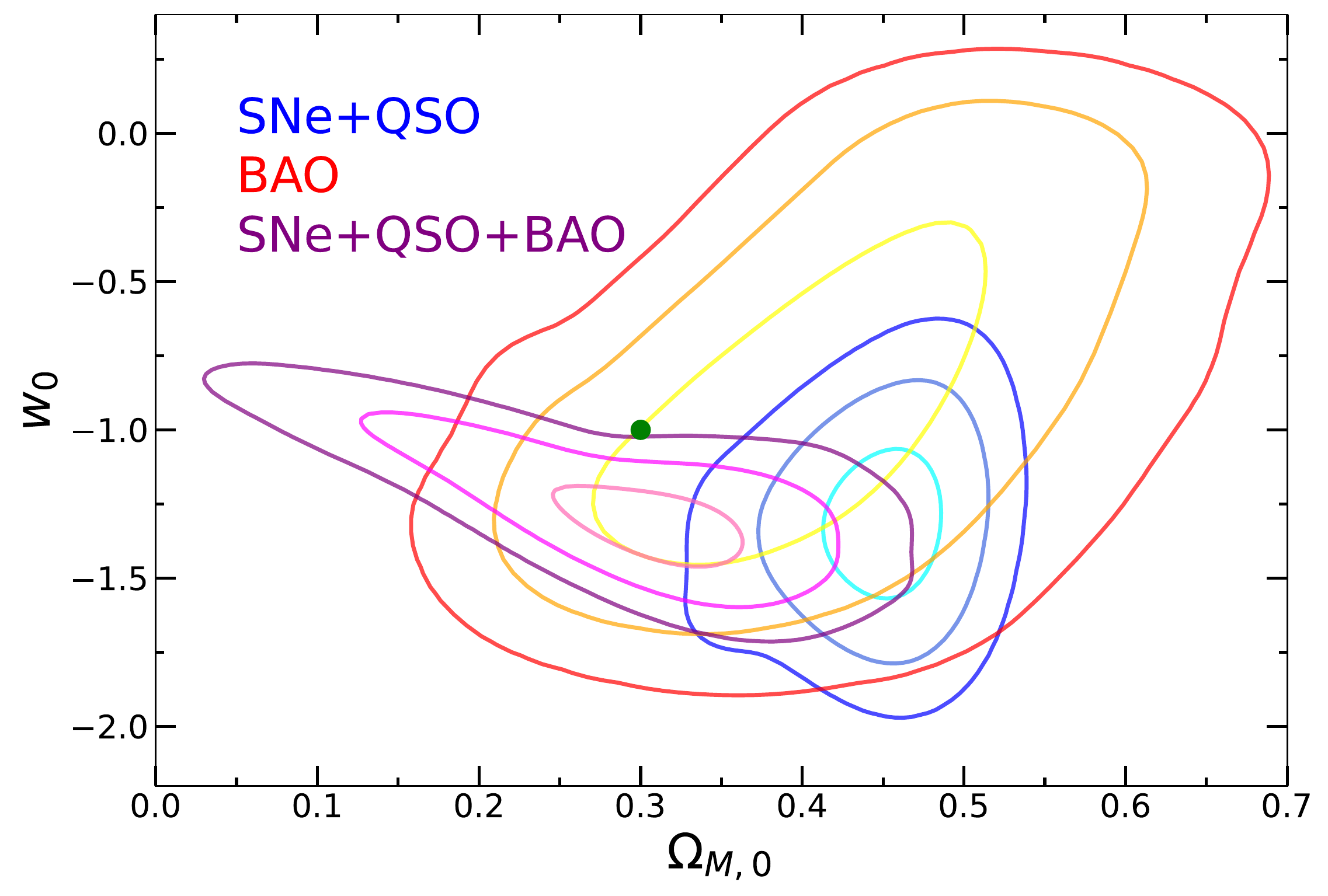}\includegraphics[width=6.2cm]{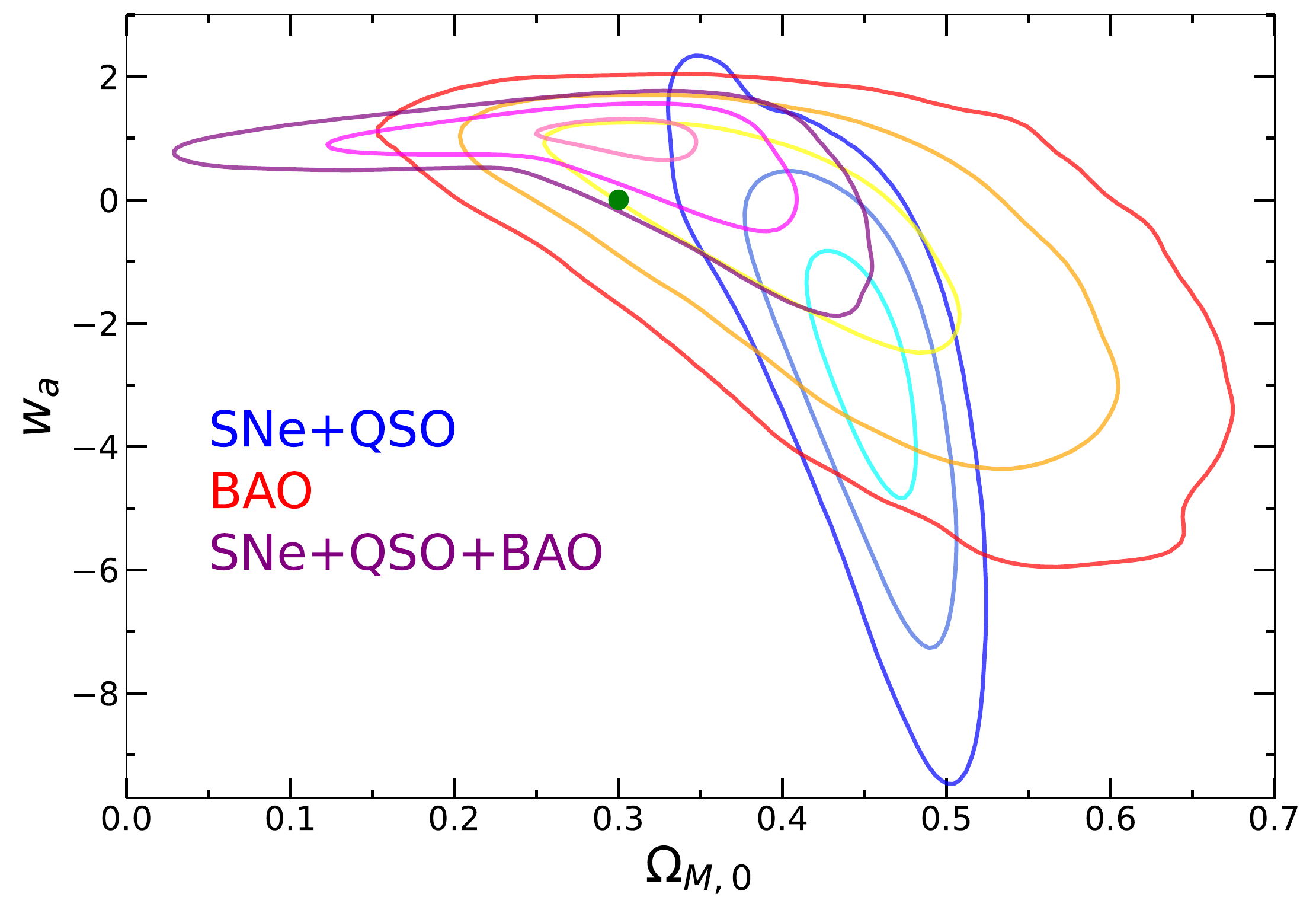}\includegraphics[width=6.2cm]{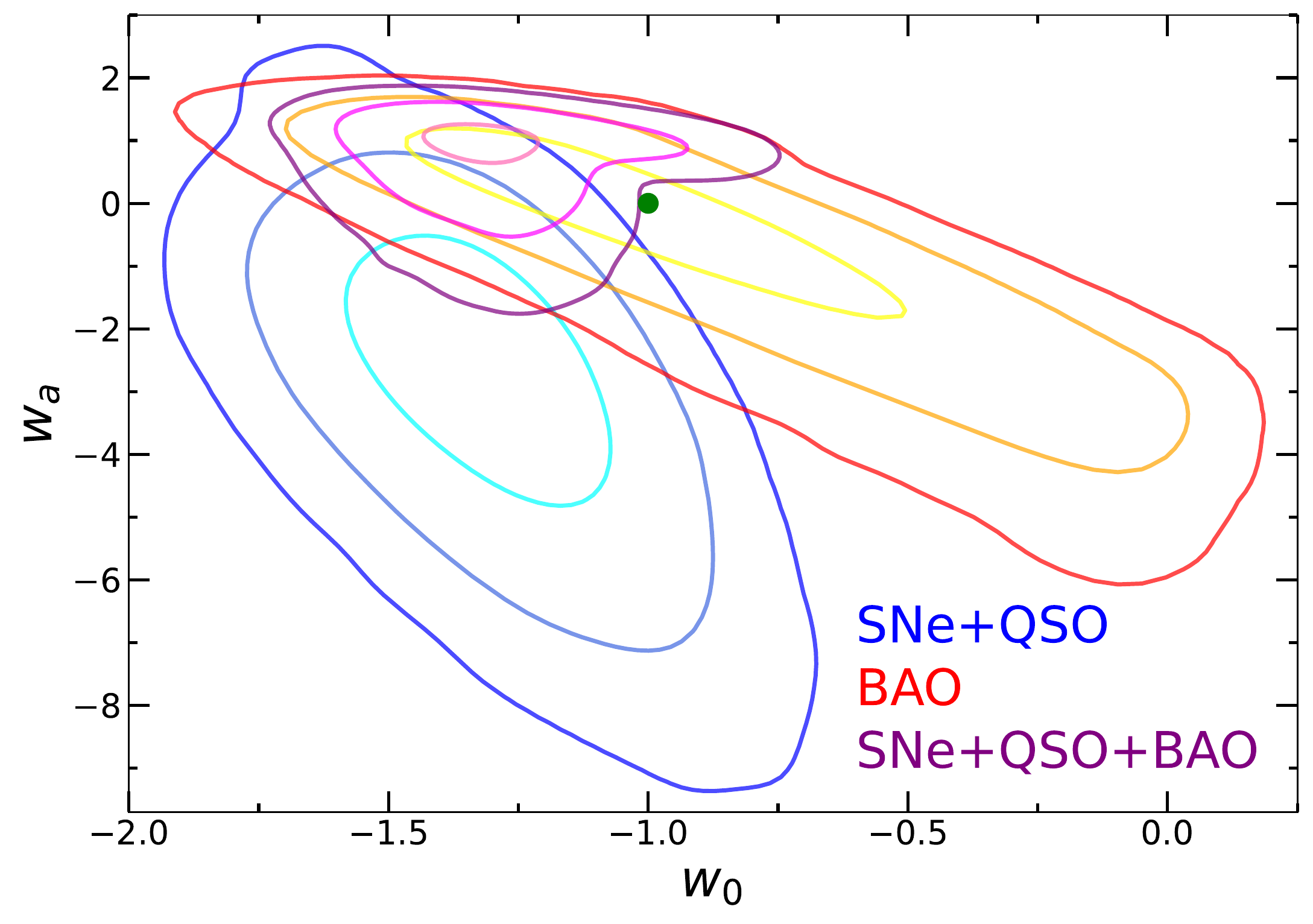}
    \caption{Bi-dimensional contour plots with 1-3$\sigma$ confidence levels obtained from each data set in the legend for the flat BA model.  The green point corresponds to a flat $\Lambda \mathrm{CDM}$ model with $\Omega_{M,0}=0.3$.} 
    \label{rationalqso+sne+bao}
\end{figure*}

\subsection{Constraints on JBP model}

Figure \ref{jbpqso+sne+bao} shows the results of this model for each pair of the cosmological free parameters. The data set SNe+QSO (blue contours) shows once again a strong discrepancy from the flat $\Lambda \mathrm{CDM}$ model with $\Omega_{M,0} = 0.3$, favouring $\Omega_{M,0}>0.3$, $w_{0}<-1$ and $w_{a} \sim -4$. The BAO alone (red contours) instead are consistent at 1$\sigma$ level with the same reference model. Despite this difference, the constraints from the two samples are compatible, also due to the large uncertainties (mainly on BAO); this allows us to fit them together. The joint fit (purple contours) results in a $2-3\sigma$ discrepancy from $\Omega_{M,0} = 0.3$, $w_{0}=-1$ and $w_{a}=0$, in the direction of a more negative $w_{0}$ and greater values of $\Omega_{M,0}$ and $w_{a}$. The shift in sign of $w_{a}$ and the correlations (or anti-correlations) between each pair of parameters are the same already seen in the CPL model. The best-fit values of all the free parameters in each data set agree with the ones obtained in the previous models.

\subsection{Constraints on exponential model}

In this model, we deal once again with only two cosmological free parameters: $\Omega_{M,0}$ and $w_{0}$. Figure \ref{expqso+sne+bao} shows the results of the fit with the same colours-data sets association used in the previous sections. Unlike the other models studied, in this case, we cannot have a direct comparison with the flat $\Lambda \mathrm{CDM}$ model in the bi-dimensional contour plot of Fig. \ref{expqso+sne+bao}. Indeed, there is no value of $w_{0}$ that turns Eq. \eqref{expmodel} in Eq. \eqref{lcdm}. This is the reason why we do not show the green point in the contour plot, as in all the other figures. Nevertheless, $\Omega_{M,0}$ and $w_{0}$ have always the same physical meanings; this means that we can compare their best-fit values with the ones expected in a flat $\Lambda \mathrm{CDM}$ model. The analysis of this model shows that the SNe+QSO sample (blue contours) prefers $\Omega_{M,0} \sim 0.4$ and $w_{0} \sim -1.5$, corresponding to a phantom DE scenario, while the BAO sample (red contours) tends to a $w_{0}$ consistent with $w_{0}=-1$, even if with a value of $\Omega_{M,0}$ always close to $\Omega_{M,0}=0.4$. The best-fit from the joint sample (purple contours) keeps the same $\Omega_{M,0}$, while $w_{0}$ is pushed to values more negative than $w_{0}=-1$. Correspondingly, the best-fit values of $\Omega_{M,0}$ and $w_{0}$ are at 2-3$\sigma$ from $\Omega_{M,0}=0.3$ and $w_{0}=-1$. These results completely agree with the ones of the flat $w$CDM model, both for the best-fit values and the correlation among parameters in each data set. The best-fit values of $\Omega_{M,0}$ and $w_{0}$ are also consistent with the ones from the CPL model. Nevertheless, the compatibility with the results of other models is always partial due to the fact that this parameterisation does not reduce to the other models for any $w_{0}$, as already explained before.

\subsection{Constraints on BA model}
\label{resultsrational}

As anticipated, to fit this model on SNe+QSO we must require a strict prior on $\Omega_{M,0}$ (i.e. $ 0.34 \leq \Omega_{M,0} \leq 1$). Otherwise, we would deal with the same convergence issues described in Sect. \ref{resultscpl} and Appendix \ref{appendixcpl}. Figure \ref{rationalqso+sne+bao} shows the bi-dimensional contour plots obtained from all data sets. Comparing once again with a flat $\Lambda \mathrm{CDM}$ model with $\Omega_{M,0} = 0.3$, the best-fits from SNe+QSO and BAO samples show respectively a tension greater than 3$\sigma$ and a consistency at 1$\sigma$ level. Nevertheless, their constraints are consistent. Their combination changes the sign of the best-fit value of $w_{a}$, as in the previous models, and shifts $\Omega_{M,0}$ towards values consistent with $\Omega_{M,0} = 0.3$ and $w_{0}$ toward values more negative than $w_{0}=-1$, with an overall tension of about 3$\sigma$ from the reference flat $\Lambda \mathrm{CDM}$ prediction. In all the data sets the best-fit values agree with the ones from the other models and the correlation (or anti-correlation) between parameters is the same already obtained in the CPL and JBP models.

\section{Summary \& Conclusions}
\label{Conclusions}
In this work, we analysed the $\Lambda \mathrm{CDM}$ and $w$CDM models in both spatially flat and non-flat assumptions, as well as some of DE extensions as the CPL and JBP parameterisations, the one we called ``exponential'' \citep[see also][]{2019PhRvD..99d3543Y}, and the BA parameterisation, using SNe Ia, QSOs, and BAO as cosmological probes. This study is strongly motivated by the need for testing the predictions of the spatially flat $\Lambda \mathrm{CDM}$ model and searching for possible deviations to explain both the theoretical and observational shortcomings of this model. The inclusion of QSOs in the cosmological analysis is crucial to this aim as they extend the Hubble diagram of SNe up to a higher-redshift range ($z=2.4-7.5$) in which predictions from different cosmological models can be distinguished and more easily compared to the observational data.
Also, we explored the compatibility of the cosmological data used, which otherwise make the results of their joint analysis misleading. It is indeed necessary to explore possible tensions between the BAO, SNe, and QSO data and their implications for the non-flat Universe and extensions of the standard cosmological model.
Below, we briefly summarise our main results.
\begin{itemize}
    \item Under the assumption of a spatially flat $\Lambda \mathrm{CDM}$ model, $\Omega_{M,0}$ is completely consistent with $\Omega_{M,0}=0.3$ in all the data sets, as expected by the latest cosmological observations \citep[e.g.][]{Hinshaw_2013,planck2018,scolnic2018}. Nevertheless, all the other models show a deviation from this prediction, with a statistical significance that is always of the order of 2-3$\sigma$ for the combined sample of SNe+QSO+BAO (Table \ref{tablefits}). We point out that the cosmological parameters obtained from the distinct samples of SNe+QSO and BAO are consistent (within 2$\sigma$) in all models and so they can be combined, except for the non-flat cases studied. In these models, indeed, BAO confirm the flatness condition while SNe+QSO show evidence of a closed Universe (Figs.~\ref{nonflatlcdmqso+sne+bao} and \ref{nonflatwcdmqso+sne+bao}).
    \item The models with a DE density evolving in time show a conclusive common trend with respect to the prediction of the $\Lambda \mathrm{CDM}$ model (Figs.~\ref{wcdmqso+sne+bao}, \ref{cplqso+sne+bao}, \ref{jbpqso+sne+bao}, \ref{expqso+sne+bao}, and \ref{rationalqso+sne+bao}). Indeed, the combination SNe+QSO+BAO always prefers $\Omega_{M,0}>0.3$, $w_{0}<-1$ and $w_{a}$ greater but consistent with $w_{a}=0$. This phantom DE behaviour is mainly driven by the contribution of SNe+QSO, while the BAO data set is generically statistically in agreement with the prediction of the flat $\Lambda \mathrm{CDM}$ model (also due to larger uncertainties) and it is responsible for the shift toward $\Omega_{M,0}=0.3$ and $w_{a}=0$. 
\end{itemize}

In conclusion, while BAO measurements are always in agreement with the standard model, the combined analyses SNe+QSO+BAO always rule out the flat $\Lambda \mathrm{CDM}$ model with $\Omega_{M,0}=0.3$, green point in the figures, at 2-3$\sigma$, with evidence of a phantom DE scenario in all the models where the DE density is allowed to vary with time. 
On the other hand, also BAO data alone are always in similar agreement with all the alternative models discussed here.
To further explore the statistical significance for such a discrepancy with the standard model, it is mandatory to improve the sample statistics at high redshifts ($z>2$), where QSOs dominate. Nonetheless, it is only with the combination of multiple perspectives and with the optimal cosmological probes at different redshifts that we can move forward to better understand the evolution of the Universe.

\section*{Acknowledgements}
GB, MB, and SC acknowledge Istituto Nazionale di Fisica Nucleare (INFN), sezione di Napoli, \textit{iniziative specifiche} QGSKY and MOONLIGHT-2. 
EL and GR acknowledge financial contribution from the agreement ASI-INAF n.2017-14-H. EL acknowledges the support of grant ID: 45780 Fondazione Cassa di Risparmio Firenze.

\section*{Data Availability}

The data underlying this article will be shared upon a reasonable request to the corresponding author.



\bibliographystyle{mnras}
\bibliography{example} 




\appendix

\section{Inclusion of SNe systematic uncertainties}
\label{appendixsys}
\begin{figure}
    \centering
    \resizebox{\hsize}{!}{\includegraphics{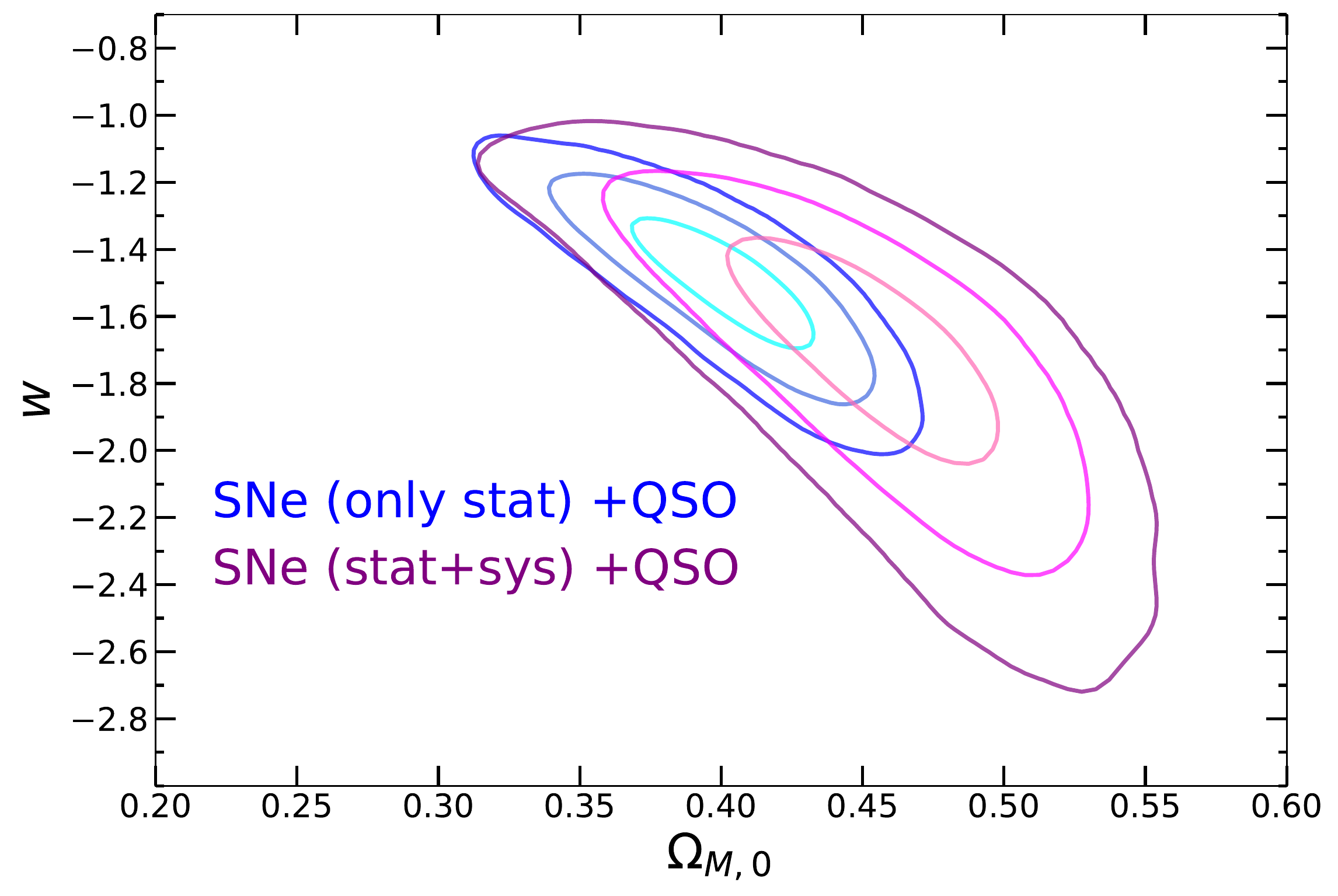}}
    \caption{Bi-dimensional contour plot with 1-3$\sigma$ confidence levels obtained from each data set in the legend for the flat $w$CDM model.} 
    \label{flatwcdmstat+sys}
\end{figure}

\begin{figure*}
    \centering
    \includegraphics[width=6.2cm]{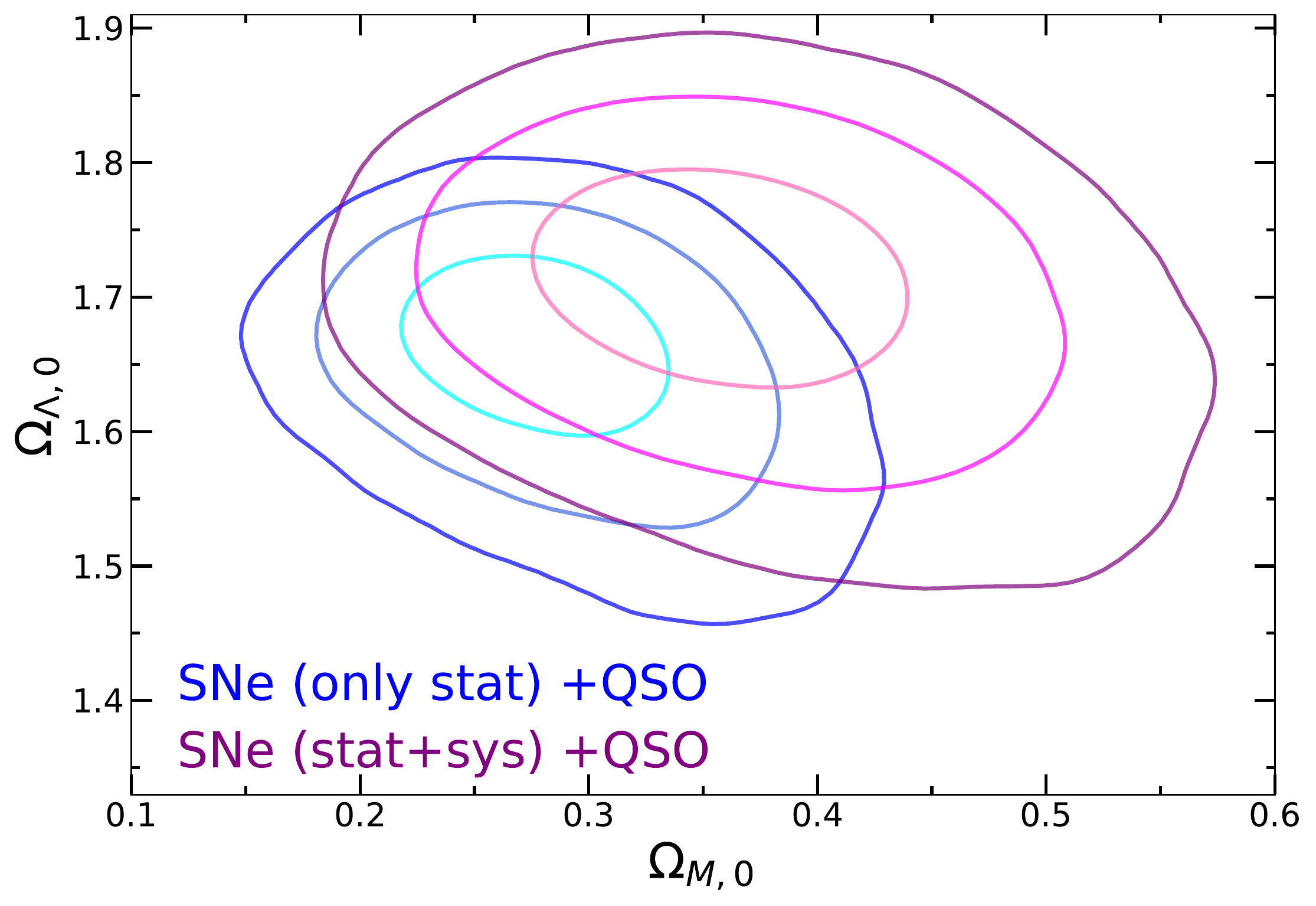}\includegraphics[width=6.2cm]{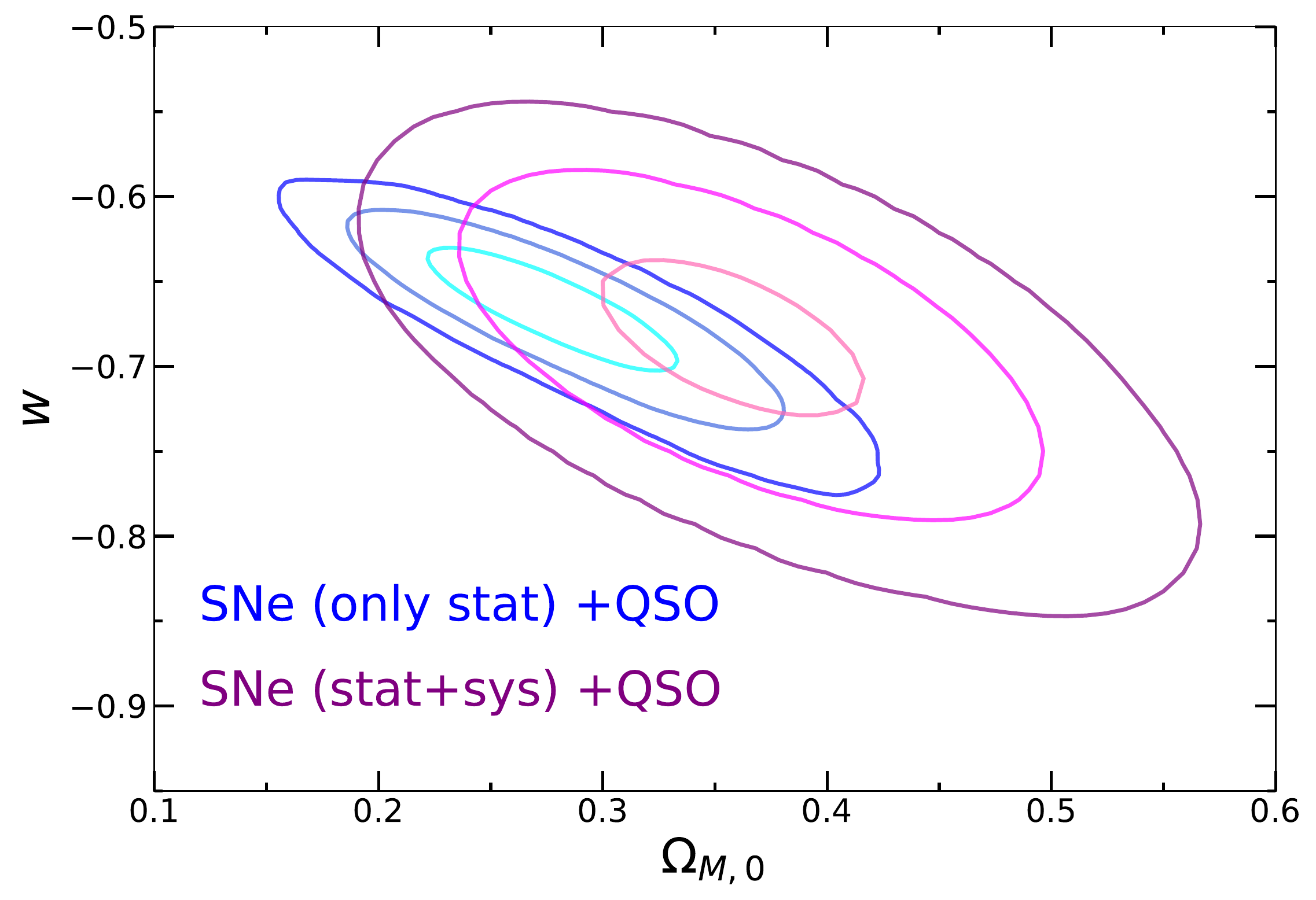}\includegraphics[width=6.2cm]{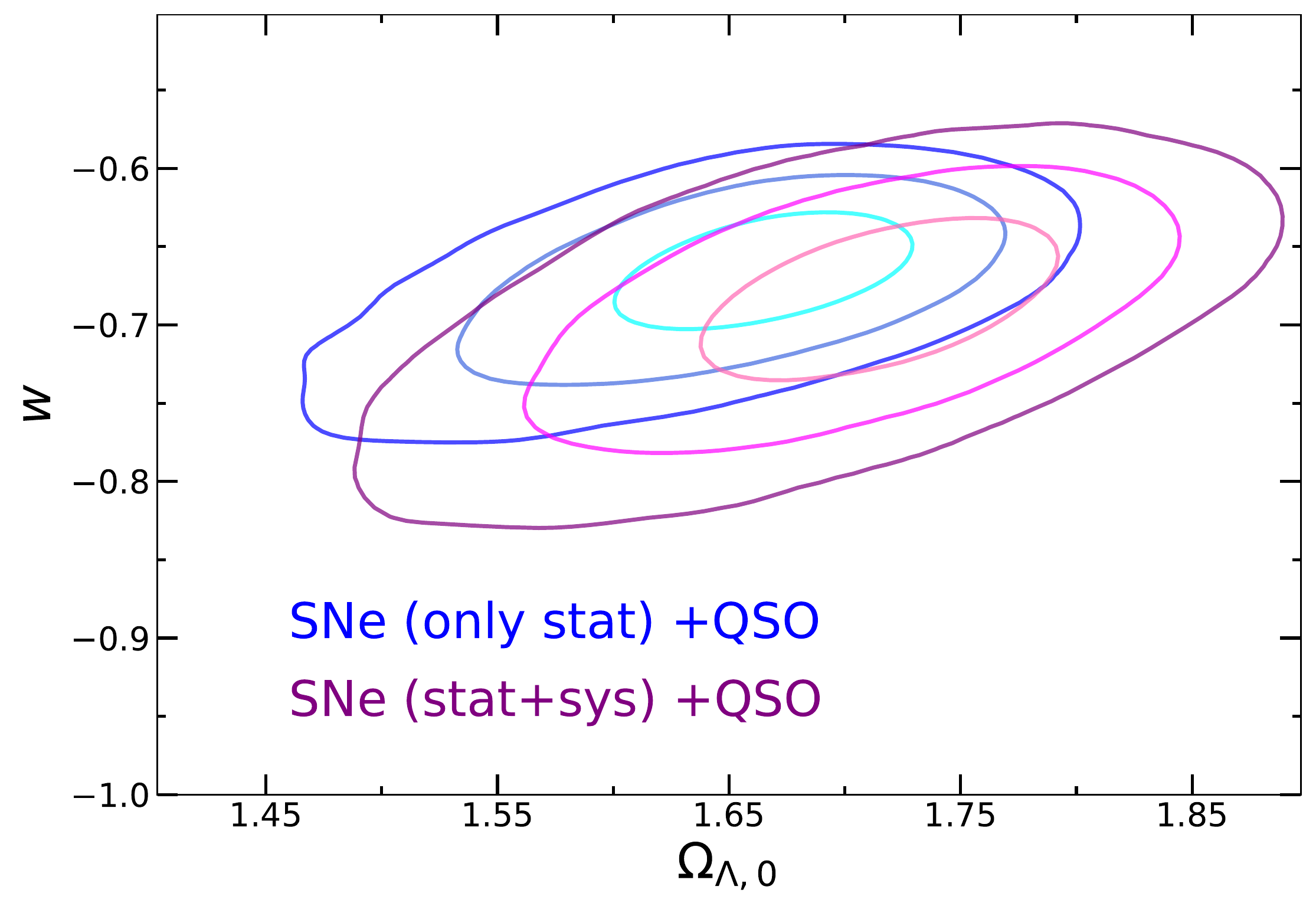}
    \caption{Contour plots for all the pairs of cosmological free parameters in the non-flat $w$CDM model for both SNe (only stat) +QSO and SNe (stat+sys) +QSO samples.} 
    \label{nonflatwcdmstat+sys}
\end{figure*}

We here show the comparison between the analyses on the SNe+QSO sample with and without the SNe systematic uncertainties. We present only the cases for the flat and non-flat $w$CDM model as all the other models lead to the same conclusions. As shown in Figs. \ref{flatwcdmstat+sys} and \ref{nonflatwcdmstat+sys}, the cosmological free parameters $\Omega_{M,0}$ and $w$ (and $\Omega_{\Lambda,0}$ for the non-flat model) are consistent within 1$\sigma$ and the correlations (or anti-correlations) remain unchanged, while the uncertainties increase in the case with the inclusion of systematics. This proves that our results do not depend on the choice on the SNe uncertainties used in the analyses.

\section{Convergence issues and extended fit of the flat CPL model}
\label{appendixcpl}

\begin{figure}
    \resizebox{\hsize}{!}{\includegraphics{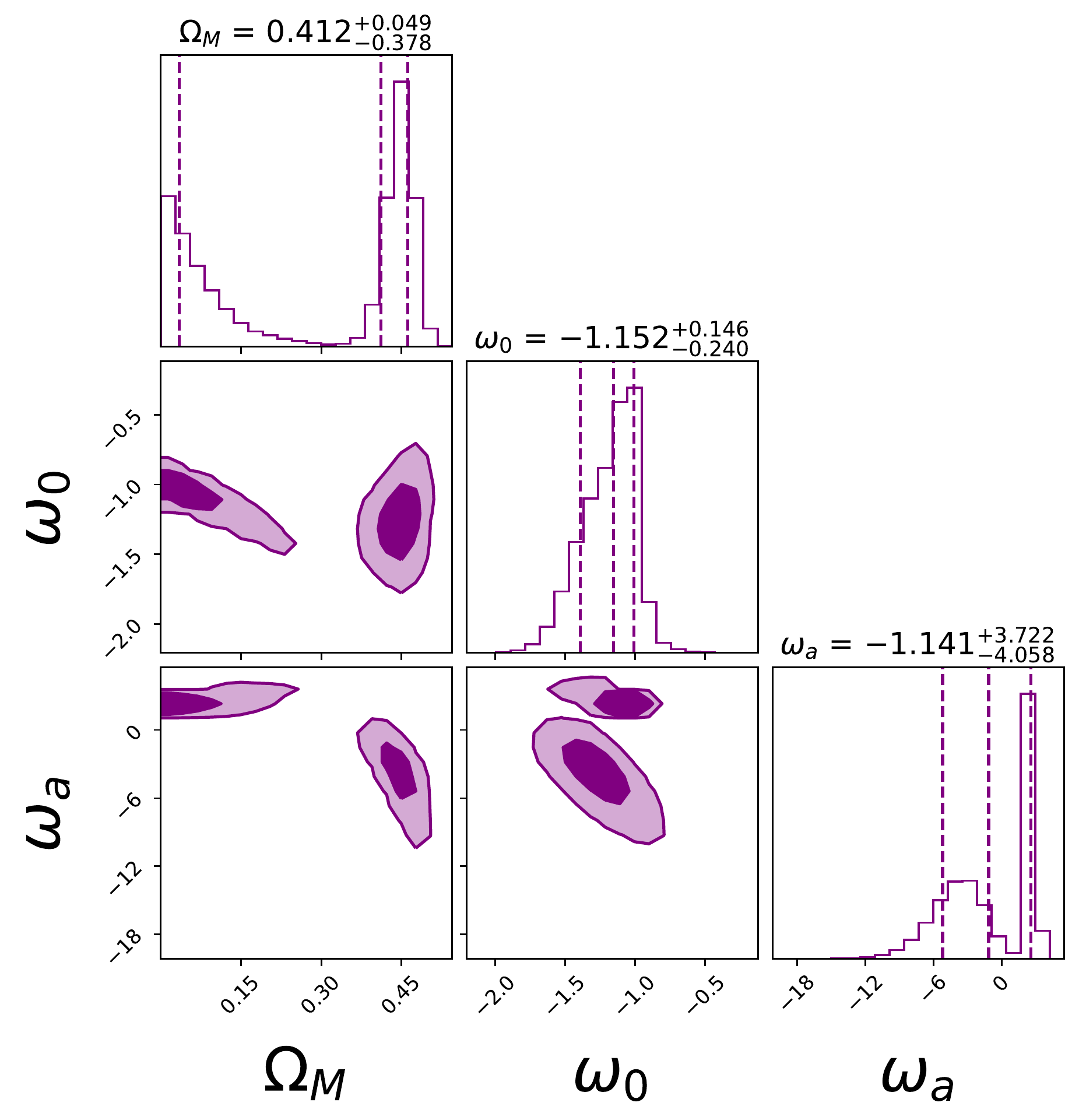}}
    \caption{Triangle plot from the fit of the CPL model on the SNe+QSO data set without any a priori requirements on the cosmological free parameters.}
    \label{2peakcpl}
\end{figure}

As stated in Sect. \ref{resultscpl}, in the CPL model we need a very strict and ad-hoc prior on $\Omega_{M,0}$ when using the SNe+QSO data set to obtain a proper convergence of the fitting algorithm toward a physical solution. For sake of clarity, here we show the result of the fit of the CPL model on the SNe+QSO sample without any specific prior. The resulting triangle plot in Fig. \ref{2peakcpl} shows two separated families of solutions: the non-physical one with a value of $\Omega_{M,0}$ that crosses $\Omega_{M,0}=0$ towards negative values and $w_{a}>0$, and the physical one with $\Omega_{M,0}>0.3$ and $w_{a}<0$. Both of them share almost the same best-fit value for $w_{0}$ close to $w_{0} \sim -1.2$, as proved by the corresponding marginalised 1D posterior distributions. All these considerations apply exactly also to the fit of the flat BA model on the SNe+QSO data set. For this reason, the triangle plot of this model does not add any useful information and we do not show it here.

\bsp	
\label{lastpage}
\end{document}